\documentclass[12pt]{iopart}
\usepackage{iopams}
\usepackage{cite}
\usepackage{array}
\usepackage[dvipsnames]{xcolor}
\usepackage{graphicx}
\usepackage[colorlinks=true,citecolor=blue]{hyperref}
\usepackage{ulem}
\usepackage{subcaption}
\usepackage{caption}
\newcolumntype{L}{>{$}l<{$}} 
\expandafter\let\csname equation*\endcsname\relax

\expandafter\let\csname endequation*\endcsname\relax

\usepackage{mathtools}
\eqnobysec

\newcommand{\subD}{\textrm{\tiny{\rm {D}}}}
\newcommand{\subS}{\textrm{\tiny{\rm {S}}}}
\newcommand{\subA}{\textrm{\tiny{\rm {A}}}}
\newcommand{\subN}{\textrm{\tiny{{\rm {N}}}}}
\newcommand{\subE}{\textrm{\tiny{{\rm {E}}}}}
\newcommand{\subR}{\textrm{\tiny{{\rm {R}}}}}
\newcommand{\subESU}{\textrm{\tiny{\rm {ESU}}}}

\begin{document}

\title[Quantum field theory on global anti-de Sitter space with Robin boundary conditions]{Quantum field theory on global anti-de Sitter space-time with Robin boundary conditions}

\author{Thomas Morley}

\address{Consortium for Fundamental Physics, School of Mathematics and Statistics,\\ Hicks Building, Hounsfield Road, Sheffield. S3 7RH United Kingdom}
\ead{TMMorley1@sheffield.ac.uk}

\author{Peter Taylor}
\address{Centre for Astrophysics and Relativity, School of Mathematical Sciences,\\ Dublin City University, Glasnevin, Dublin 9, Ireland}
\ead{Peter.Taylor@dcu.ie}

\author{Elizabeth Winstanley}
\address{Consortium for Fundamental Physics, School of Mathematics and Statistics,\\ Hicks Building, Hounsfield Road, Sheffield. S3 7RH United Kingdom}
\ead{E.Winstanley@sheffield.ac.uk}

\begin{abstract}
We compute the vacuum polarization for a massless, conformally coupled scalar field on the covering space of global, four-dimensional, anti-de Sitter space-time. 
Since anti-de Sitter space is not globally hyperbolic, boundary conditions must be applied to the scalar field.
We consider general Robin (mixed) boundary conditions for which the classical evolution of the field is well-defined and stable.
The vacuum expectation value of the square of the field is not constant unless either Dirichlet or Neumann boundary conditions are applied.
We also compute the thermal expectation value of the square of the field.
For Dirichlet boundary conditions, both thermal and vacuum expectation values approach the same well-known limit on the space-time boundary.
For all other Robin boundary conditions (including Neumann boundary conditions), the vacuum and thermal expectation values have the same limit on the space-time boundary, but this limit does not equal that in the Dirichlet case.
\end{abstract}


\vspace{2pc}

\section{Introduction}
\label{sec:intro}
Quantum field theory (QFT) on anti-de Sitter (AdS) space-time has been the subject of considerable attention owing to its role in the holographic principle and string theory, particularly within the context of the AdS/CFT (conformal field theory) correspondence (see for example \cite{Aharony:1999ti} for a review).
QFT on AdS is particularly rich, with a plethora of possibilities to consider.
As on any space-time, one can study a variety of bosonic 
\cite{Allen:1986ty,Allen:1985wd,Ambrus:2018olh,Avis:1977yn,Barroso:2019cwp,Belokogne:2016dvd,Burgess:1984ti,Caldarelli:1998wk,Camporesi:1991nw,Camporesi:1992wn,Dappiaggi:2016fwc,Dappiaggi:2017wvj,Dappiaggi:2018pju,Dappiaggi:2018xvw,Kent:2014nya} 
and fermionic  
\cite{Allen:1986qj,Ambrus:2018olh,Ambrus:2014fka,Ambrus:2015mfa,Ambrus:2017cow,Ambrus:2017vlf,Camporesi:1992tm,Camporesi:1992wn,Cotaescu:2007xv,Muck:1999mh}
quantum fields,
and different quantum states, including static vacuum states 
\cite{Ambrus:2018olh,Ambrus:2015mfa,Barroso:2019cwp,Belokogne:2016dvd,Caldarelli:1998wk,Camporesi:1992wn,Kent:2014nya},
static thermal states
\cite{Allen:1986ty,Ambrus:2018olh,Ambrus:2014fka,Ambrus:2017cow,Ambrus:2017vlf}
and rotating states
\cite{Ambrus:2014fka,Kent:2014wda}.

Let us now consider the simplest possible quantum field, namely a massless, conformally coupled scalar field.
Even in this simplified model, there are many variations to consider.
First of all, the properties of the QFT of the scalar field depend on whether one considers global AdS
\cite{Allen:1986ty,Ambrus:2018olh,Avis:1977yn,Barroso:2019cwp,Caldarelli:1998wk,Camporesi:1992wn,Dappiaggi:2018xvw,Kent:2014nya}
or the Poincar\'e patch PAdS
\cite{Dappiaggi:2016fwc,Dappiaggi:2017wvj,Dappiaggi:2018pju,Pitelli:2019svx,Pitelli:2019pua,Pitelli:2019xeg},
the latter being particularly relevant in the context of the AdS/CFT correspondence.
In both cases, the fact that AdS is not a globally hyperbolic space-time means that, in order to have a well-defined QFT, appropriate boundary conditions must be applied to the field at null infinity, which is a time-like surface
\cite{Avis:1977yn,Benini:2017dfw,Dappiaggi:2017wvj,Dappiaggi:2018pju,Dappiaggi:2018xvw,Ishibashi:2003jd,Ishibashi:2004wx,Wald:1980jn}.

The simplest boundary conditions are either Dirichlet 
\cite{Allen:1986ty,Ambrus:2018olh,Avis:1977yn,Kent:2014nya}  (where the field vanishes on the boundary) or Neumann \cite{Allen:1986ty,Avis:1977yn} (where the normal derivative of the field vanishes on the boundary). In \cite{Avis:1977yn} a third possibility is also studied, namely ``transparent'' boundary conditions, which we do not consider further in this paper. 
The advantage of working with either Dirichlet or Neumann boundary conditions is that the vacuum Green's function respects the maximal symmetry of the background AdS space-time \cite{Allen:1985wd,Avis:1977yn}, which enables renormalized vacuum expectation values to be derived in closed form \cite{Allen:1986ty,Ambrus:2018olh,Kent:2014nya} and analytic expressions for renormalized thermal expectation values can be found in terms of infinite sums of special functions \cite{Allen:1986ty,Ambrus:2018olh}.

However, Dirichlet and Neumann boundary conditions are not the only possibilities leading to well-defined dynamics for a classical scalar field \cite{Dappiaggi:2020yxg}.  
For example, one can also consider Robin (or mixed) boundary conditions, in which a linear combination of the field and its normal derivative vanish on the boundary \cite{Ishibashi:2004wx,Gannot:2018jkg}, or Wentzell boundary conditions \cite{Dappiaggi:2018pju}.
In the AdS/CFT correspondence, Robin boundary conditions for a bulk quantum scalar field have been extensively studied (see 
\cite{Berkooz:2002ug,Casper:2017gcw,DelGrosso:2019gow,Gubser:2002zh,Minces:2002wp,Minces:2004zr,Minces:1999eg,Minces:2001zy,Nolland:2003kc,Vecchi:2010dd} for an incomplete selection of references on this topic), and correspond to multi-trace deformations of the dual CFT.

In this paper, we focus on the role the boundary conditions play for both vacuum and thermal states of a massless, conformally coupled scalar field on the covering space of global AdS in four space-time dimensions. 
Some of the interesting questions arising in this context are:
\begin{enumerate}
\item \label{qu1} Are general Robin boundary conditions physically valid?
    \item \label{qu2} Are vacuum and thermal states Hadamard for general Robin boundary conditions?
    \item \label{qu3} Do the propagators for vacuum and thermal states respect the AdS symmetries for all Robin boundary conditions?
    \item \label{qu4} Practically, how does one efficiently compute quantum expectation values for both vacuum and thermal states for arbitrary Robin boundary conditions?
    \item \label{qu5} Do quantum expectation values such as the vacuum polarization asymptote to a finite value for arbitrary Robin boundary conditions?
\end{enumerate}
There are at least partial answers to some of these questions scattered throughout the literature
\cite{Barroso:2019cwp,Dappiaggi:2016fwc,Dappiaggi:2018xvw,Ishibashi:2004wx,Pitelli:2019svx,Pitelli:2019xeg}. 
In answer to (\ref{qu1}), consistent dynamics for a classical scalar field can be formulated for a subset of Robin boundary conditions \cite{Ishibashi:2004wx}.
Hadamard ground states for the quantum scalar field can be constructed for at least some Robin boundary conditions \cite{Dappiaggi:2016fwc,Dappiaggi:2018xvw}, partially answering (\ref{qu2}), although these states are no longer maximally symmetric \cite{Barroso:2019cwp,Pitelli:2019svx} (\ref{qu3}).
Recently, the study of (\ref{qu4}, \ref{qu5}) has been initiated with a computation of the renormalized vacuum polarization and stress-energy tensor for a massless, conformally coupled scalar field for which most of the field modes satisfy Dirichlet boundary conditions, but the s-wave modes satisfy Robin boundary conditions \cite{Barroso:2019cwp}.
It is found that the expectation values are not maximally symmetric but they asymptote to their values for Dirichlet boundary conditions as the space-time boundary is approached. 

Any attempt to answer (\ref{qu1}--\ref{qu5}) concretely for general scalar field mass, coupling and numbers of space-time dimensions is rather complicated, and several different cases need to be considered \cite{Ishibashi:2004wx,Dappiaggi:2018xvw}. 
In this paper we therefore restrict our attention to four space-time dimensions and a massless, conformally coupled scalar field in order to simplify both presentation and computations, and enable the underlying features to be discerned. 
We also consider global AdS rather than PAdS.
In the latter case there exist bound state modes \cite{Troost:2003ig} which render the construction of ground states more involved \cite{Dappiaggi:2016fwc}, but these bound state modes are absent on global AdS \cite{Dappiaggi:2018xvw,Minces:2004zr}.
Our focus in this paper is addressing points (\ref{qu4}--\ref{qu5}).
We consider the simplest possible expectation value, the vacuum polarization (square of the field).
We develop a methodology which enables the efficient computation of this quantity for Robin boundary conditions, and employ this to 
present novel results for the vacuum polarization for conformal scalar fields for which {\em {all}} modes satisfy general Robin boundary conditions.

We begin, in section~\ref{sec:classical}, with a brief review of the classical mode solutions of the Klein-Gordon equation for a massless, conformally coupled scalar field on AdS, before turning to the canonical quantization of the field in section~\ref{sec:Lorentzian}.
We derive a mode-sum expression for the Wightman function for the vacuum state, with Robin boundary conditions applied consistently to all field modes.
This expression does not lend itself to a practical method of computing renormalized expectation values, 
so in section~\ref{sec:Euclidean} we consider the related problem of constructing thermal states on the Euclidean section of AdS. 
We obtain a mode-sum representation of the Euclidean Green's function for both vacuum and thermal states, again with Robin boundary conditions applied to all field modes.
From this we are able to readily compute the renormalized vacuum polarization for both thermal and vacuum states when Robin boundary conditions are applied. 
Our conclusions are presented in section~\ref{sec:conc}.

\section{Classical scalar field on CAdS}
\label{sec:classical}

AdS is a maximally symmetric solution of the Einstein equations with a negative cosmological constant $\Lambda=-3/L^{2}$, where $L$ is the AdS curvature length-scale related to the Ricci scalar by $R=-12/L^{2}$ in four space-time dimensions. 
In global coordinates, the AdS metric is given by
\begin{equation}
\label{eq:AdS}
    ds^{2}=L^{2}\sec^{2}\rho\left(-\rmd t^{2}+\rmd\rho^{2}+\sin^{2}\rho \,\rmd\Omega_{2}^{2}\right),
\end{equation}
where $t\in(-\pi,\pi]$ with the end-points identified, $\rho\in [0,\pi/2)$ and $\rmd\Omega_{2}^{2}$ is the line-element for the two-sphere $\mathbb{S}^{2}$.
The periodicity of the time coordinate implies the existence of closed time-like curves, a problem that is circumvented by ``unwrapping'' the time coordinate.
This defines the covering space of AdS (hereafter denoted by CAdS) which has the same line-element as (\ref{eq:AdS}) but with $t\in (-\infty,\infty)$. 
Even in the covering space, the space-time is not globally hyperbolic. 
In particular the boundary $\rho=\pi/2$ is time-like and it is necessary to impose boundary conditions here in order to define the field theory
\cite{Avis:1977yn,Benini:2017dfw,Dappiaggi:2017wvj,Dappiaggi:2018pju,Dappiaggi:2018xvw,Ishibashi:2003jd,Ishibashi:2004wx,Wald:1980jn}. 
This requirement has a significant impact on the QFT.

Specializing to a conformally invariant scalar field, one can use the fact that CAdS is conformal to half of the Einstein Static Universe (ESU) to impose boundary conditions on fields in the latter space-time. 
Letting $g_{\mu\nu}$ be the metric components of CAdS and $\widetilde{g}_{\mu\nu}$ the components of the ESU metric in these coordinates, we then have
\begin{equation}
    \widetilde{g}_{\mu\nu}=\Omega^{2}\,g_{\mu\nu},\qquad \Omega=\cos\rho.
    \label{eq:conformaltransformation}
\end{equation}
CAdS is thus conformal to the portion of the ESU for which $\rho \in [0, \pi/2)$, which is half of the full ESU space-time \cite{Avis:1977yn}.

\subsection{Scalar field modes}
\label{sec:modes}

The wave equation for the conformal scalar field on ESU is
\begin{equation}
   \left\{ \widetilde{\Box}-\frac{1}{6}\widetilde{R}\right\}
   \widetilde{\varphi}(x)=\left\{ \widetilde{\Box}-\frac{1}{L^{2}}\right\}\widetilde{\varphi}(x)=0,
   \label{eq:ESUwave}
\end{equation}
where all quantities with a tilde are with respect to the ESU metric $\widetilde{g}_{\mu\nu}$. 
A complete set of solutions of this equation is given by
\begin{equation}
    \widetilde{\varphi}_{\omega \ell m}(x)\sim \rme^{-\rmi \omega t}Y^{m}_{\ell}(\theta,\phi)\widetilde{\chi}_{\omega \ell}(\rho),
    \label{eq:ESUmodes}
\end{equation}
where $Y^{m}_{\ell}(\theta,\phi)$ with $\ell\in \mathbb{N}$ and $m=-\ell,...,\ell$ are the spherical harmonics and $\widetilde{\chi}_{\omega \ell}(\rho)$ satisfies the radial equation
\begin{equation}
   \left\{ \frac{\rmd}{\rmd \rho}\left(\sin^{2}\rho\frac{\rmd}{\rmd\rho}\right)
-(1-\omega^{2})\sin^{2}\rho-\ell(\ell+1)\right\}\widetilde{\chi}_{\omega \ell}(\rho)=0.
\label{eq:chieqn}
\end{equation}
The general solution of (\ref{eq:chieqn}) is
\begin{equation}
\widetilde{\chi}_{\omega \ell}(\rho)=(\sin\rho)^{-1/2}\left[C_{1}P_{\omega-\frac{1}{2}}^{\ell+\frac{1}{2}}(\cos\rho)+C_{2}Q_{\omega-\frac{1}{2}}^{\ell+\frac{1}{2}}(\cos\rho)\right],
\label{eq:chigen}
\end{equation}
where $P^{\mu}_{\nu}(z)$, $Q^{\mu}_{\nu}(z)$ are associated Legendre functions and $C_{1}$, $C_{2}$ are arbitrary constants.
Demanding that the solution be regular at the origin requires $C_{1}=0$.
In general $Q^{\mu}_{\nu}(z)$ is ill-defined whenever $\nu+\mu$ is a negative integer. 
Therefore we employ Olver's definition of the Legendre function of the second kind \cite{NIST:DLMF}
\begin{equation}
    \mathbf{Q}^{\ell+1/2}_{\omega-1/2}(\cos\rho)=\frac{Q^{\ell+1/2}_{\omega-1/2}(\cos\rho)}{\Gamma(\omega+\ell+1)},
\end{equation}
which is valid for all $\ell$ and $\omega$. 
However, whenever $\omega$ is an integer such that $\omega\le \ell$, then $\mathbf{Q}_{\omega-1/2}^{\ell+1/2}(\cos\rho)=0$. 
We can, without loss of generality, set $C_{2}=1$ in (\ref{eq:chigen}) since the overall constant is set by the normalization of the mode solutions. 
Hence, we take
\begin{equation}
    \widetilde{\chi}_{\omega \ell}(\rho)=(\sin \rho)^{-1/2}\mathbf{Q}^{\ell+1/2}_{\omega-1/2}(\cos\rho).
\end{equation}

Now we impose boundary conditions at the timelike boundary at $\rho=\pi/2$ in CAdS by imposing boundary conditions on $\widetilde{\chi}_{\omega \ell}(\rho)$ at $\rho=\pi/2$. 
We can parametrize the general Robin boundary conditions by an angle $\alpha\in [0,\pi )$ so that
\begin{equation}
\label{eq:Robin}
    \widetilde{\chi}_{\omega \ell}(\rho) \cos \alpha 
    +\frac{\rmd\widetilde{\chi}_{\omega \ell}(\rho)}{\rmd \rho}
    \sin \alpha =0,\quad \rho\to \pi/2.
\end{equation}
With this parametrization, Dirichlet boundary conditions correspond to $\alpha=0$ while Neumann boundary conditions correspond to $\alpha=
\pi/2$.
We shall assume that the parameter $\alpha  $ is a constant, although more general boundary conditions for which $\alpha $ is not constant also lead to a well-defined initial/boundary value problem for the scalar field \cite{Gannot:2018jkg}.

Equation~(\ref{eq:Robin}) leads to the following quantization condition on the mode frequency $\omega$,
\begin{equation}
\label{eq:RobinQuantization}
-\tan\left(\frac{1}{2}[\ell+\omega]\pi\right)\frac{\Gamma(\frac{\omega-\ell}{2})\Gamma(\frac{\omega+\ell+1}{2})}{\Gamma(\frac{\omega+\ell+2}{2})\Gamma(\frac{\omega-\ell+1}{2})}=2\tan\alpha.
\end{equation}
For each $\ell $, there is a discrete set of quantized frequencies satisfying (\ref{eq:RobinQuantization}). 
We denote these frequencies as $\omega _{n\ell }$, where $n=1,2,\ldots $ indexes the solutions of (\ref{eq:RobinQuantization}) for each fixed $\ell $.
While giving an explicit expression for the discrete set of frequencies $\omega_{n\ell}$ that solve the transcendental equation (\ref{eq:RobinQuantization}) is impossible in general, for Dirichlet boundary conditions we find $\omega_{n\ell}=2n-\ell$ and for Neumann boundary conditions $\omega_{n\ell}=2n+1-\ell$. 
Recall that for $\omega$ an integer, we must have $\omega>\ell$, which for the Dirichlet case means $n>\ell$ and for the Neumann case $n\ge \ell$.

\begin{figure}
    \centering
    \includegraphics[width=.8\linewidth]{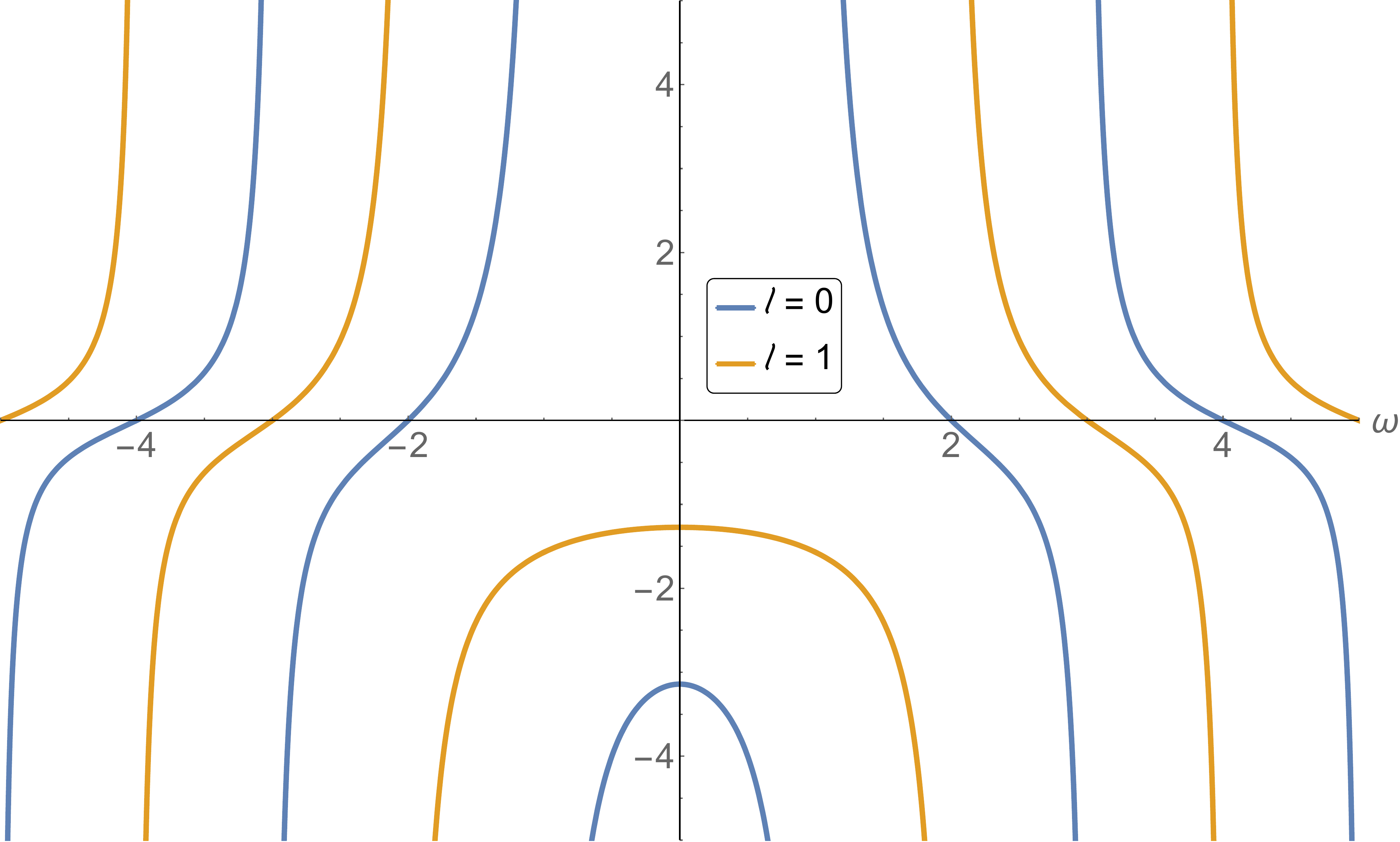}
    \caption{Left-hand-side of the quantization condition (\ref{eq:RobinQuantization}) for $\ell = 0$ and $1$, as a function of the frequency $\omega $.}
    \label{fig:quantization}
\end{figure}

For general $\alpha $, the quantized frequencies $\omega _{n\ell }$ satisfying (\ref{eq:RobinQuantization}) will not be integers.
For fixed $\ell $, the left-hand-side of (\ref{eq:RobinQuantization}) vanishes when $\ell + \omega $ is an even integer and diverges when $\ell + \omega $ is an odd integer, taking all real values for $\ell + 2n-1\le \omega \le \ell+ 2n+1$, with $n=1,2, \ldots $ (see Figure~\ref{fig:quantization}).
Therefore there is a unique solution $\omega _{n\ell }$ to the quantization condition (\ref{eq:RobinQuantization}) in each interval $\ell + 2n-1\le \omega \le \ell+ 2n+1$, with $n=1,2, \ldots $.
For $0 < \omega <  \ell + 1$, the left-hand-side of (\ref{eq:RobinQuantization}) is negative and has a maximum at $\omega =0$, where it is greater than or equal to $-\pi $.
If $\tan \alpha < - \frac{\pi }{2}$, there is therefore an additional solution to (\ref{eq:RobinQuantization}) in the interval $0<\omega < \ell + 1$. 

Finally, the mode solutions $\varphi_{n\ell m}(x)$ of the scalar wave equation on CAdS are simply obtained from (\ref{eq:ESUmodes}) by using the
conformal transformation (\ref{eq:conformaltransformation}). This gives $\varphi _{n\ell m}=\widetilde{\varphi }_{n\ell m} \cos\rho $, and hence
\begin{equation}
\label{eq:modesolutions}
    \varphi_{n\ell m}(x)=N_{n\ell}\rme^{-\rmi \omega_{n\ell} t}Y^{m}_{\ell}(\theta,\phi)\cos\rho\,(\sin\rho)^{-1/2}\mathbf{Q}^{\ell+1/2}_{\omega_{n\ell}-1/2}(\cos\rho),
\end{equation}
where $\omega_{n\ell}$ satisfies the quantization condition (\ref{eq:RobinQuantization}) and $N_{n\ell}$ is a normalization constant.

\subsection{Classical instabilities}
\label{sec:instability}

A massless, conformally coupled scalar field satisfies the Breitenlohner-Freedman bound \cite{Breitenlohner:1982jf,Breitenlohner:1982bm},
which implies that the field is stable when either Dirichlet or Neumann boundary conditions are imposed.
Ishibashi and Wald \cite{Ishibashi:2004wx} have proven the more general result that, for any real value of the parameter $\alpha $ governing the Robin boundary conditions (\ref{eq:Robin}), the dynamics of the classical scalar field are well-defined, in other words, $\alpha $ labels a one-parameter family of self-adjoint extensions ${\mathfrak {A}}_{\alpha }$ of the radial differential operator ${\mathfrak {A}}$ governing the field.
However, these self-adjoint extensions are not necessarily positive.
For values of $\alpha $ for which ${\mathfrak {A}}_{\alpha }$ fails to be a positive operator, the dynamics of the field will be unstable, in the sense that generic perturbations will grow unboundedly in time.
This instability will be manifest in the existence of mode solutions of the scalar field equation having imaginary frequency.

Setting $\omega = \rmi \Omega $ with $\Omega $ real, the quantization condition (\ref{eq:RobinQuantization}) becomes
\begin{equation}
\label{eq:UnstableModes}
    -\frac{|\Gamma(\frac{\rmi \Omega +\ell+1}{2})|^{2}}{|\Gamma(\frac{\rmi \Omega +\ell+2}{2})|^{2}}
    = 2\tan\alpha ,
\end{equation}
where we have simplified using properties of the $\Gamma $ function \cite{NIST:DLMF}. 
If $\alpha = 0$ (Dirichlet boundary conditions) or $\alpha = \frac{\pi }{2}$ (Neumann boundary conditions), equation (\ref{eq:UnstableModes}) has no solutions and there are no unstable modes, in line with the results described above.
Furthermore, it is clear from (\ref{eq:UnstableModes}) that there are no unstable modes if $0<\alpha <\frac{\pi }{2}$.

\begin{figure}
    \centering
    \includegraphics[width=.8\linewidth]{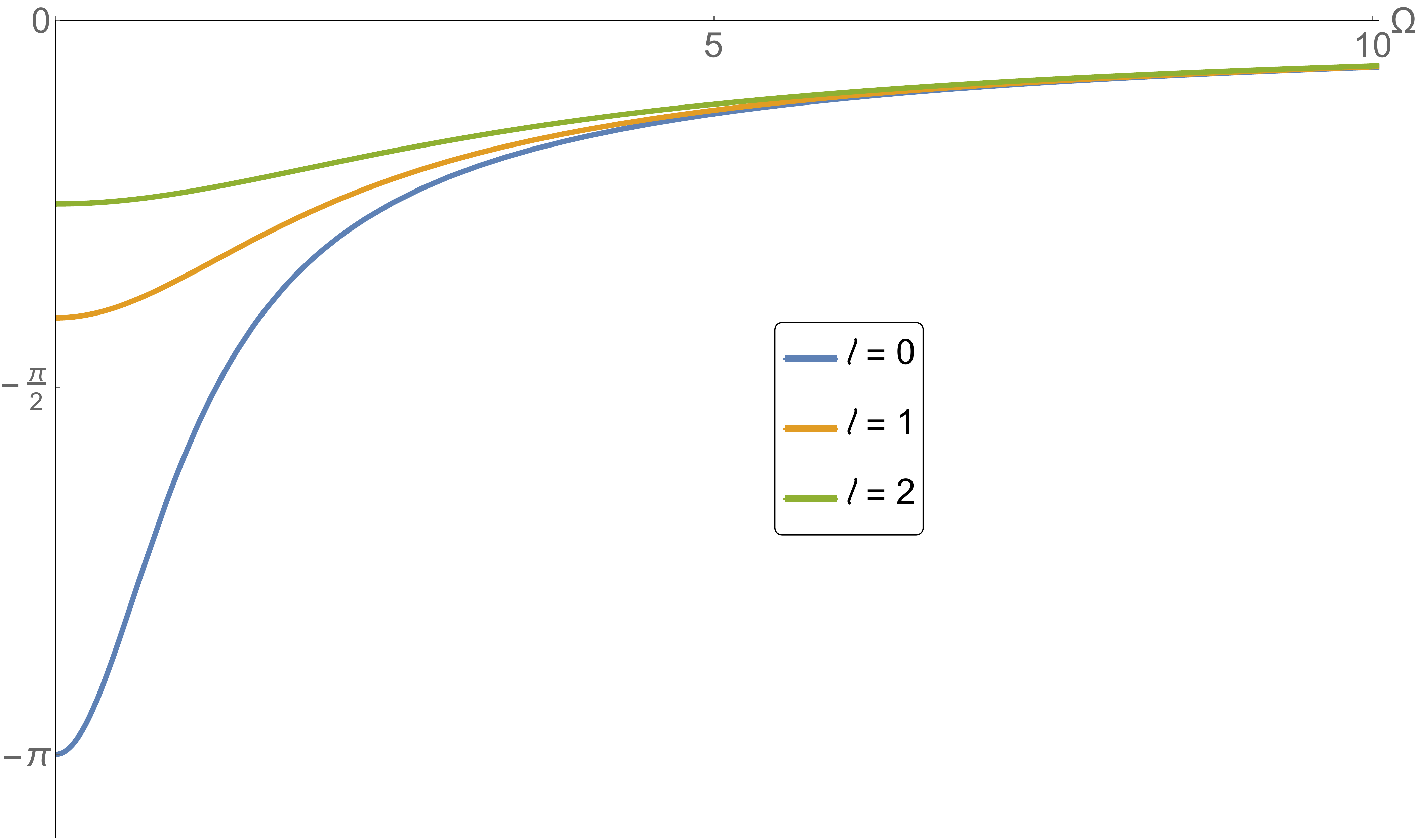}
    \caption{Left-hand-side of the instability condition (\ref{eq:UnstableModes}) for $\ell = 0$, $1$ and $2$, as a function of the frequency $\Omega $.}
    \label{fig:instability}
\end{figure}

However, unstable modes exist for some $\alpha $ such that $\tan \alpha <0$.
Using the asymptotic properties of the $\Gamma $ functions \cite{NIST:DLMF},
it can be proven that the supremum of the left-hand-side of (\ref{eq:UnstableModes}) is zero.
Furthermore, for fixed $\Omega $, the left-hand-side of (\ref{eq:UnstableModes}) is an increasing function of $\ell$, while for fixed $\ell$, it is symmetric in $\Omega $ and increasing as $\Omega >0$ increases, with minimum value $-\pi $ (see Figure \ref{fig:instability}).
Therefore there exist real $\Omega $ satisfying (\ref{eq:UnstableModes}) if $-\frac{\pi }{2} < \tan \alpha < 0$, which corresponds to $-\tan ^{-1} \left( \frac{\pi }{2} \right) < \alpha <0$, or, equivalently, $\pi /2 < \alpha <\pi -\tan ^{-1} \left( \frac{\pi }{2} \right)$, where $\tan ^{-1}\left( \frac{\pi }{2}\right) \approx 0.32\pi $ and $\pi -\tan ^{-1}\left( \frac{\pi }{2}\right) \approx 0.68 \pi $.
We can only consider a quantum scalar field for values of $\alpha $ for which the classical set-up is stable, so for the rest of this paper we restrict our attention to $\alpha\in  \left[0,\alpha _{\rm {crit}}\right]$,
where 
\begin{equation}
    \alpha _{\rm {crit}} = \pi -\tan ^{-1}\left( \frac{\pi }{2}\right) \approx 0.68 \pi .
    \label{eq:alphacrit}
\end{equation}

\section{Quantum scalar field on CAdS}
\label{sec:Lorentzian}

We now describe the canonical quantization of the massless conformally coupled scalar field on four-dimensional Lorentzian CAdS space-time.
The Wightman function for vacuum states with Robin boundary conditions is constructed in section~\ref{sec:canonical} (see also \cite{Dappiaggi:2018xvw}).
This two-point function is divergent in the limit in which the points are brought together, and the regularization of this divergence is discussed in section~\ref{sec:regularization}.
Next we compute the renormalized vacuum polarization for Dirichlet and Neumann boundary conditions in section~\ref{sec:VPDN}, and validate our approach by rederiving the well-known results for these boundary conditions  \cite{Allen:1986ty}.
In section~\ref{sec:VPRobinL} we discuss the practical difficulties inherent in the computation for Robin boundary conditions.

\subsection{Canonical quantization}
\label{sec:canonical}

The standard procedure for quantizing a classical scalar field is to promote the field to an operator-valued distribution $\varphi(x)\to\hat{\varphi}(x)$ and then to impose canonical commutation relations on this operator (see, for example, \cite{Birrell:1982ix}). 
These commutation relations imply that the following two-point function,
\begin{equation}
    G_{A}(x,x')=\rmi \langle A|\mathcal{T}\left\{\hat{\varphi}(x),\hat{\varphi}(x')\right\}|A\rangle
    \label{eq:Feynman}
\end{equation}
is in fact a Green's function for the scalar wave operator, the so-called Feynman propagator for the field in the state $|A\rangle$. 
The operator $\mathcal{T}$ appearing in this definition is a time-ordering operator given by
\begin{equation}
  \mathcal{T}\{\hat{\varphi}(t,\textbf{x}), \hat{\varphi}(t',\textbf{x}')\} = \begin{cases} 
      \hat{\varphi}(t,\textbf{x}) \hat{\varphi}(t',\textbf{x}') & \mbox{if $t>t'$}, \\
      \hat{\varphi}(t',\textbf{x}') \hat{\varphi}(t,\textbf{x}) & \mbox{if $t'>t$},
      \end{cases}
\end{equation}
and $|A\rangle$ is assumed to be a unit-norm quantum state. 
We will find it convenient to express the Feynman Green's function in terms of the Wightman two-point function 
\begin{equation}
    G^{+}_{A}(x,x')=\left[ G^{-}_{A}(x,x')\right]^{\dagger}=\langle A|\hat{\varphi}(x)\hat{\varphi}(x')|A\rangle ,
\end{equation}
which is related to the Feynman propagator by
\begin{equation}
G_{A}(x,x')=\rmi \,\Theta(t-t')\,G^{+}_{A}(x,x')+\rmi \,\Theta(t'-t)\,G^{-}_{A}(x,x') ,
\end{equation}
where $\Theta(z)$ is the step function. 

In this section we focus on vacuum states, which we will denote by $|0\rangle_{\alpha}$, making explicit the dependence on the parameter $\alpha $ governing the boundary condition (\ref{eq:Robin}). 
Since we have a globally static coordinate system, a natural vacuum is defined by expanding the quantum field in a basis of positive frequency modes with respect to our time coordinate $t$. 
The Dirichlet and Neumann boundary conditions combined with regularity at the origin already enforced that the frequency be positive in those cases, since we have $\omega_{n\ell}$ an integer such that $\omega_{n\ell}>\ell$. 
More generally, we can express the Wightman function for the vacuum states for arbitrary boundary conditions by \cite{Dappiaggi:2018xvw}
\begin{equation}
\label{eq:Wightman}
    G^{+}_{\alpha}(x,x')=\sum_{\ell=0}^{\infty}\sum_{m=-\ell}^{\ell}\sum_{\omega_{n\ell}}\varphi_{n\ell m}(x)\varphi_{n\ell m}^{*}(x'),
\end{equation}
where $\varphi_{n\ell m}(x)$ are given by (\ref{eq:modesolutions}). We note that the sum over frequencies must be performed first since these depend on $\ell$. 

It remains to compute the normalization constant $N_{n\ell}$ appearing in (\ref{eq:modesolutions}). 
In fact, implicit in the expression (\ref{eq:Wightman}) is the assumption that $\varphi_{n\ell m}(x)$ are orthonormal with respect to an appropriate inner product. 
In a globally hyperbolic space-time, the inner product $\left<\varphi_{1}, \varphi_{2}\right>$ of any two solutions $\varphi _{1}$, $\varphi _{2}$ of the scalar field equation is taken to be
\begin{equation}
\label{eq:KGproduct}
\left<\varphi_{1}, \varphi_{2}\right>=\rmi \int_{\Sigma}(\varphi_{1}^{*}\partial_{\mu}\varphi_{2}-\varphi_{2}\partial_{\mu}\varphi_{1}^{*}) n^{\mu} \, \rmd\Sigma
\end{equation}
where $\Sigma$ is any Cauchy surface and the integral is independent of the choice of Cauchy surface. 
In CAdS, we must also specify data on the boundary at $\rho=\pi/2$ (which we denote ${\mathcal {I}}_{\pi /2}$). We require that the inner product is independent of the choice of space-like hypersurface $\Sigma$ and also independent of the boundary conditions imposed on the solutions. 

\begin{figure}
    \centering
    \includegraphics[width=.6\linewidth]{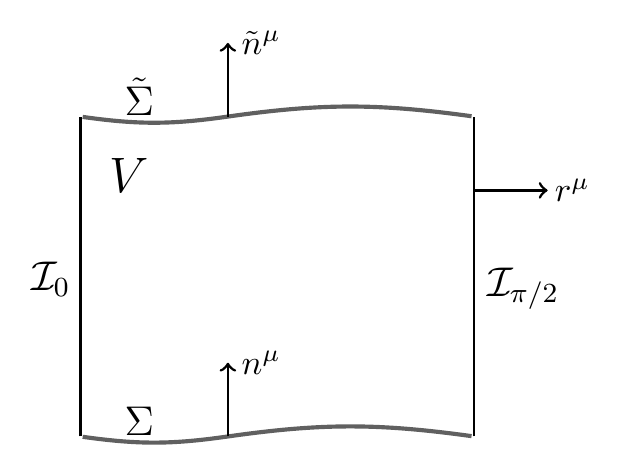}
    \caption{Diagram showing the volume $V$ bounded by $S=\mathcal{I}_{0}\cup\mathcal{I}_{\pi/2}\cup\Sigma\cup\widetilde{\Sigma}$ for the application of Stokes' theorem.}
    \label{fig:Stokes}
\end{figure}

To see this, let $V$ be the volume region delimited by the boundary $S=\mathcal{I}_{0}\cup\mathcal{I}_{\pi/2}\cup\Sigma\cup\widetilde{\Sigma}$ where $\mathcal{I}_{0}$ is the time-like hypersurface defined by $\rho=0$, while $\Sigma$ and $\widetilde{\Sigma}$ are space-like hypersurfaces with unit future-pointing normals $n^{\mu}$ and $\widetilde{n}^{\mu}$, respectively (see Figure~\ref{fig:Stokes}). 
Using Stokes' Theorem, we have
\begin{equation}
\int_{S}\left(\varphi_{1}^{*}\partial_{\mu}\varphi_{2}-\varphi_{2}\partial_{\mu}\varphi_{1}^{*}\right) \, \rmd S^{\mu}=
\int_{V}\nabla^{\mu}\left(\varphi_{1}^{*}\partial_{\mu}\varphi_{2}-\varphi_{2}\partial_{\mu}\varphi_{1}^{*}\right) \, \rmd V.
\label{eq:stokes}
\end{equation}
The right-hand-side vanishes on account of the scalar field equation. 
The left-hand-side can be written as a sum of the contributions from each of the boundary terms. 
We can show that the contribution from $\mathcal{I}_{0}$ vanishes by noting that
\begin{equation}
    \chi_{\omega\ell}(\rho)=\frac{(-1)^{\ell+1}\pi}{2^{\ell+3/2}\Gamma(\ell+3/2)\Gamma(\omega-\ell)}\rho^{\ell}+\Or (\rho^{\ell+2}).
\end{equation}
It is clear that all $\ell>0$ modes vanish at $\rho=0$, while the derivative of the $\ell=0$ mode vanishes at $\rho =0$.  
Combining these implies that the integrand is zero on $\mathcal{I}_{0}$. 
Putting these together, we require
\begin{eqnarray}
    0 & = & 
    \int_{\mathcal{I}_{\pi/2}}\left(\varphi_{1}^{*}\partial_{\mu}\varphi_{2}-\varphi_{2}\partial_{\mu}\varphi_{1}^{*}\right)r^{\mu} \, \rmd\mathcal{I}
    +\int_{\widetilde{\Sigma}}\left(\varphi_{1}^{*}\partial_{\mu}\varphi_{2}-\varphi_{2}\partial_{\mu}\varphi_{1}^{*}\right)\widetilde{n}^{\mu} \, \rmd \widetilde{ \Sigma}\nonumber\\ & & 
    -\int_{\Sigma}\left(\varphi_{1}^{*}\partial_{\mu}\varphi_{2}-\varphi_{2}\partial_{\mu}\varphi_{1}^{*}\right)n^{\mu} \, \rmd \Sigma
\end{eqnarray}
where $r^{\mu}$ is the outward pointing normal to the timelike boundary $\rho=\pi/2$. 
The minus sign on the last term is a result of the fact that we have defined both $n^{\mu}$ and $\widetilde{n}^{\mu}$ to be future-pointing. 
Given $\varphi_{1}$, $\varphi_{2}$ satisfying general Robin boundary conditions
(\ref{eq:Robin}), the boundary conditions themselves immediately imply
\begin{equation}
    \varphi_{1}^{*}\partial_{\mu}\varphi_{2}=\varphi_{2}\partial_{\mu}\varphi_{1}^{*}=-\left(\tan\rho+\cot\alpha\right)\varphi_{1}^{*}\varphi_{2},
\end{equation}
and hence the contribution to the surface integral on the boundary also vanishes. 
We are therefore  left with
\begin{equation}
    \int_{\widetilde{\Sigma}}\left(\varphi_{1}^{*}\partial_{\mu}\varphi_{2}-\varphi_{2}\partial_{\mu}\varphi_{1}^{*}\right)\widetilde{n}^{\mu}
    \, \rmd \widetilde{\Sigma}=\int_{\Sigma}\left(\varphi_{1}^{*}\partial_{\mu}\varphi_{2}-\varphi_{2}\partial_{\mu}\varphi_{1}^{*}\right)n^{\mu} \, \rmd \Sigma.
\end{equation}
Therefore the inner product (\ref{eq:KGproduct}) with $\Sigma$ an arbitrary space-like hypersurface is independent of the choice of hypersurface and the Robin boundary conditions applied. 

Equipped with a suitable inner product, the normalization constant $N_{n\ell }$ is determined by insisting the modes are orthonormal:
\begin{equation}
    \langle \varphi_{n\ell m}(x), \varphi_{n'\ell' m'}(x)\rangle=\delta_{n n'}\delta_{\ell \ell'}\delta_{m m'}.
\end{equation}
After applying the orthonormality of the spherical harmonics, we obtain
\begin{equation}
    \langle \varphi_{n\ell m}(x), \varphi_{n'\ell' m'}(x)\rangle=\delta_{\ell\ell'}\delta_{m m'}L^{2}(\omega_{n\ell}+\omega_{n'\ell})N_{n\ell}N_{n'\ell}\int_{0}^{\pi/2}\tan^{2}\rho\,\chi_{n\ell}(\rho)\chi_{n'\ell}(\rho) \, \rmd\rho.
\end{equation}
The integral here can be performed, but is rather tedious so we relegate the calculation to the appendix where it is shown that
\begin{eqnarray}
\fl    \int_{0}^{\pi/2}\tan^{2}\rho\,\chi_{n\ell}(\rho)\chi_{n'\ell}(\rho) \, \rmd\rho & = & 
\delta_{n n'}\frac{\pi\left[\pi-\sin(\pi(\omega_{n\ell}+\ell))\{ \zeta(\ell+\omega_{n\ell}+1)
+\zeta(\omega_{n\ell}-\ell)\} \right]}{8\,\omega_{n\ell}\,\Gamma(\ell+\omega_{n\ell}+1)\Gamma(\omega_{n\ell}-\ell)},
\nonumber \\ & &
\label{eq:intappendix}
\end{eqnarray}
where
\begin{equation}\label{eq:zeta}
    \zeta(z)=\frac{1}{2}\left[ \psi\left(\frac{z+1}{2}\right)-\psi\left(\frac{z}{2}\right)\right].
\end{equation}
Therefore the normalization constant is
\begin{equation}
\label{eq:norm}
    N_{n\ell}^{2}=\frac{4\,\Gamma(\ell+\omega_{n\ell}+1)\Gamma(\omega_{n\ell}-\ell)}{L^{2}\pi\left[\pi-\sin(\pi(\omega_{n\ell}+\ell))\{\zeta(\ell+\omega_{n\ell}+1)+\zeta(\omega_{n\ell}-\ell) \} \right]}.
\end{equation}

The Wightman function (\ref{eq:Wightman}) can now be expressed as \cite{Dappiaggi:2018xvw}
\begin{eqnarray}
\label{eq:Wightman1}
 \fl G_{\alpha}^{+}(x,x')&=&
\frac{1}{\pi^{2}L^{2}}\frac{\cos\rho\,\cos\rho'}{\sqrt{\sin\rho\,\sin\rho'}}\sum_{\ell=0}^{\infty}(2\ell+1)P_{\ell}(\cos\gamma)\sum_{\omega_{n\ell}}\rme^{-\rmi \omega_{n\ell}\Delta t}\nonumber\\
\fl &&
\quad\times \frac{\Gamma(\ell+\omega_{n\ell}+1)\Gamma(\omega_{n\ell}-\ell)
\mathbf{Q}_{\omega_{n\ell}-1/2}^{\ell+1/2}(\cos\rho)\mathbf{Q}_{\omega_{n\ell}-1/2}^{\ell+1/2}(\cos\rho')}{\left[\pi-\sin(\pi(\omega_{n\ell}+\ell))\{ \zeta(\ell+\omega_{n\ell}+1)+\zeta(\omega_{n\ell}-\ell)\}\right]},
\end{eqnarray}
where we have used a standard addition theorem for spherical harmonics to perform the sum over $m$-modes, $P_{\ell}(x)$ are Legendre polynomials and
\begin{equation}
 \cos\gamma=\cos\theta\cos\theta'+\sin\theta\sin\theta'\cos\Delta\phi  .
 \label{eq:cosgamma}
\end{equation} 
Here and throughout, we use $\Delta x=x-x'$ as a shorthand for the coordinate separation. 
In (\ref{eq:Wightman1}) and the following analysis, for compactness we have omitted an $\epsilon \rightarrow 0$ term in the exponential which is required to regulate the sum over frequencies.
This will be discussed further in section \ref{sec:VPRobinL}.

To see how the Wightman function (\ref{eq:Wightman1}) simplifies for Dirichlet boundary conditions with $\alpha=0$, we recall that the quantization condition (\ref{eq:RobinQuantization}) is in this case satisfied by  $\omega_{n\ell}=2n-\ell$ with $n>\ell$. 
As well as the normalization constant (\ref{eq:norm}) simplifying greatly, the sum over frequencies can now be given explicitly to yield
\begin{eqnarray}
\label{eq:WightmanDirichlet}
\fl G_{\subD}^{+}(x,x')&=&\frac{1}{\pi^{3}L^{2}}\frac{\cos\rho\,\cos\rho'}{\sqrt{\sin\rho\,\sin\rho'}}\sum_{\ell=0}^{\infty}(2\ell+1)P_{\ell}(\cos\gamma)\sum_{n=\ell}^{\infty}\rme^{-\rmi(2n-\ell)\Delta t}\nonumber\\
\fl &&\quad\times \Gamma(2n+1)\Gamma(2n-2\ell)\mathbf{Q}_{2n-\ell-1/2}^{\ell+1/2}(\cos\rho)\mathbf{Q}_{2n-\ell-1/2}^{\ell+1/2}(\cos\rho').
\end{eqnarray}
Similarly, for Neumann boundary conditions with $\alpha= \pi/2$, we obtain
\begin{eqnarray}
\label{eq:WightmanNeumann}
\fl G^{+}_{\subN}(x,x')&=&\frac{1}{\pi^{3}L^{2}}\frac{\cos\rho\,\cos\rho'}{\sqrt{\sin\rho\,\sin\rho'}}\sum_{\ell=0}^{\infty}(2\ell+1)P_{\ell}(\cos\gamma)\sum_{n=\ell}^{\infty}\rme^{-\rmi(2n-\ell+1)\Delta t}\nonumber\\
\fl &&\quad\times \Gamma(2n+1)\Gamma(2n-2\ell+2)\mathbf{Q}_{2n-\ell+1/2}^{\ell+1/2}(\cos\rho)\mathbf{Q}_{2n-\ell+1/2}^{\ell+1/2}(\cos\rho').
\end{eqnarray}
Note that in both of these cases, since the frequencies are integers, the two-point function is periodic in time even though we are working on the covering space CAdS. 
This implies that the vacuum states for Dirichlet and Neumann boundary conditions in CAdS are the same as those in AdS. 
There are no other values of $\alpha $ for which the quantization condition (\ref{eq:RobinQuantization}) admits integer-frequency solutions. 
Hence the equivalence between vacuum states on AdS and its covering space only holds for Dirichlet and Neumann boundary conditions.

\subsection{Regularization of quantum expectation values}
\label{sec:regularization}

In QFT in curved space-time a central role is played by expectation values of time-ordered products of the quantum fields at a particular point. 
For example, the source term in the semi-classical Einstein equations is the expectation value of the quantum stress-energy tensor operator.
Since the quantum fields are operator-valued distributions, the expectation values of such objects involve products of distributions at a given space-time point and are not mathematically well-defined. 
Therefore a regularization prescription is required to make sense of the theory.
We describe here the simplest case of regularizing the time-ordered Wick square of the quantum scalar field, which defines the so-called vacuum polarization,
\begin{equation}
\label{eq:VPDef}
\langle A|\hat{\varphi}^{2}(x)|A\rangle=-\rmi\lim_{x'\to x}\left(G_{\subA}(x,x')-G_{\subS}(x,x')\right)
\end{equation}
where $G_{\subA}(x,x')$ is the Feynman Green's function (\ref{eq:Feynman}) for the scalar field in the state $|A\rangle$ and $G_{\subS}(x,x')$ is a two-point function required to render the limit finite. 
We restrict attention to a class of quantum states that satisfy the so-called Hadamard condition, that is, states for which the Feynman Green's function has the following short-distance behaviour \cite{Decanini:2005eg}
\begin{eqnarray}
\label{eq:HadamardCondition}
\fl    G_{\subA}(x,x')=&\frac{\rmi}{8\pi^{2}}\Bigg\{\frac{U(x,x')}{(\sigma(x,x')+\rmi\epsilon)}+V(x,x')\log\left(\frac{2\sigma(x,x')}{d^{2}}+\rmi\epsilon\right)+W_{\subA}(x,x')\Bigg\}
\end{eqnarray}
where $U$, $V$ and $W_{\subA}$ are symmetric biscalars and $\sigma (x,x')$ is Synge's world function, corresponding to half the square of the geodetic distance between  the two points (assuming there is a unique geodesic connecting them). 
The parameter $d$ in (\ref{eq:HadamardCondition}) is an arbitrary length-scale needed to make the argument of the log term dimensionless. 
The term involving $U(x,x')$ above is called the direct part of the Hadamard form while the term involving $V(x,x')$ is known as the tail of the Hadamard form. 
For massless, conformally coupled scalar fields ($m=0$, $\xi=1/6$) in CAdS, $V\equiv 0$ \cite{Decanini:2005eg,Kent:2014nya}. 
Both of these terms contain all the short-distance (or ultraviolet) divergences. They are constructed only from the geometry through the metric and its derivatives. 
The remaining term $W_{\subA}(x,x')$ depends on the quantum state and cannot be determined by a local expansion.

In order to obtain a finite limit in (\ref{eq:VPDef}), we adopt what is known as the Hadamard regularization prescription \cite{Decanini:2005eg}, which simply involves taking $G_{\subS}(x,x')$ to be any symmetric locally-constructed Hadamard parametrix for the Klein-Gordon wave operator, for example, taking $G_{\subS}(x,x')$ to be
\begin{eqnarray}
\label{eq:GS}
\fl    G_{\subS}(x,x')=&\frac{\rmi}{8\pi^{2}}\Bigg\{\frac{U(x,x')}{(\sigma(x,x')+\rmi \epsilon)}+V(x,x')\log\left(\frac{2\sigma(x,x')}{d^{2}}+\rmi\epsilon\right)+W(x,x')\Bigg\},
\end{eqnarray}
where $W(x,x')$ is any regular symmetric biscalar constructed only from the geometry. The simplest choice is the trivial one $W(x,x')\equiv 0$. Making this choice, we have
\begin{eqnarray}
    \langle A|\hat{\varphi}^{2}(x)|A\rangle=\frac{1}{8\pi^{2}}w_{\subA}(x),
\end{eqnarray}
where $w_{\subA}(x)=\lim_{x'\to x}W_{\subA}(x,x')$, which is manifestly finite.

This formalism relied explicitly on the assumption that the quantum state we considered satisfied the Hadamard condition (\ref{eq:HadamardCondition}) and indeed there is general consensus that physically reasonable quantum states must be Hadamard (see for example \cite{Fewster:2013lqa}). 
A natural question then is whether the vacuum states we consider in this section here are Hadamard for all Robin boundary conditions. 
This question is addressed in \cite{Dappiaggi:2018xvw} (see also the comments in \cite{Pitelli:2019svx}) where it is shown that the vacuum states are indeed Hadamard states for all Robin boundary conditions. 
However, we add the caveat that, as discussed in section~\ref{sec:instability}, there are Robin boundary conditions for which the classical scalar field is unstable \cite{Ishibashi:2004wx} and for such values of $\alpha$, it does not make sense to consider the quantization of the scalar field. Moreover, the propagators derived above are not the correct representation of the propagator for the classically unstable scalar fields. 
Henceforth, we shall not consider Robin boundary conditions for which the field is unstable, in which case the propagators above are indeed the correct representation and so the only divergences in the propagator occur at the vertex of the lightcone (the coincidence limit) and are those contained in the Hadamard parametrix (\ref{eq:GS}) (after an appropriate `$\rmi\epsilon$' prescription has been implemented). We shall also consider mixed thermal states in what follows and the same caveat applies to these states.

\subsection{Vacuum polarization for Dirichlet and Neumann boundary conditions}
\label{sec:VPDN}

As a check of the general formalism outlined in section~\ref{sec:regularization}, we will next compute the vacuum polarization for the scalar field in the vacuum state, with Dirichlet or Neumann boundary conditions applied. 
The answer is already well-known \cite{Allen:1986ty} and follows by assuming that the Green's function for the conformal field depends only on the world function $\sigma$. 
This ansatz allows one to write down the Green function in closed form satisfying either boundary condition \cite{Avis:1977yn}. 
However, the propagator for the vacuum state for the field satisfying general Robin boundary conditions will not be maximally symmetric and a closed-form representation of the propagator will not be attainable. In those cases, implementing the regularization prescription is more subtle, a topic which we will discuss in detail in the next section.

For now, we verify that our mode-sum representation of the Wightman function (\ref{eq:WightmanDirichlet}, \ref{eq:WightmanNeumann}) yields the same answer for the vacuum polarization as that given in \cite{Allen:1986ty} using the closed-form expression.
We make use of the fact that the Dirichlet and Neumann vacuum states are maximally symmetric and hence the choice of origin is irrelevant. 
Therefore computing the vacuum polarization at $\rho=0$ will give the correct answer on the entire space-time. 
The asymptotics of the Legendre functions \cite{NIST:DLMF},
\begin{equation}
\frac{\mathbf{Q}^{\ell+1/2}_{\omega-1/2}(\cos\rho)}{\sqrt{\sin\rho}}\sim\frac{(-1)^{\ell+1}\pi\,\sin^{\ell}\rho}{2^{\ell+3/2}\Gamma(\ell+3/2)\Gamma(\omega-\ell)},\quad\rho\to 0,
\end{equation}
implies that near the origin only the $\ell=0$ mode contributes \cite{Burgess:1984ti}. 
Moreover, the $\ell=0$ radial modes for the Dirichlet case are
\begin{equation}
\fl\frac{\mathbf{Q}^{1/2}_{2n-1/2}(\cos\rho)}{\sqrt{\sin\rho}}=-\sqrt{\frac{\pi}{2}}\frac{1}{\Gamma(2n+1)}\frac{\sin (2n\rho)}{\sin\rho}\to-\sqrt{\frac{\pi}{2}}\frac{1}{\Gamma(2n)},\quad \textrm{as}\,\,\rho\to 0.
\end{equation}
Hence we have, at the origin,
\begin{equation}
G_{\subD}^{+}(\rho=\rho'=0;\Delta t)=\frac{1}{2\pi^{2}L^{2}}\sum_{n=1}^{\infty} 2n  \, \rme^{-2n\rmi\Delta t}.
\end{equation}
This sum is not convergent in the usual sense, but with an appropriate `$\rmi \epsilon$' prescription, we obtain
\begin{equation}
\label{eq:GDiepsilon}
    G_{\subD}^{+}(\rho=\rho'=0;\Delta t)=-\frac{1}{4\pi^{2}L^{2}}\lim_{\epsilon\to 0^{+}}\frac{1}{\sin^{2}(\Delta t-\rmi \epsilon)}=-\frac{1}{4\pi^{2}L^{2}}\lim_{\epsilon\to 0^{+}}\frac{1}{(\sin^{2}\Delta t-\rmi \epsilon)},
\end{equation}
where the last equality follows by absorbing a factor of $2\sin\Delta t\,\cos\Delta t$ into a redefinition of $\epsilon$ (assuming $\Delta t$ is such that $2\sin\Delta t\,\cos\Delta t>0$) and ignoring $\Or(\epsilon^{2})$ terms. From the distributional identity
\begin{equation}
\lim_{\epsilon\to 0^{+}}\frac{1}{(z^{2}-\rmi\epsilon)}=\mathcal{P}\left(\frac{1}{z^{2}}\right)+\pi\rmi\delta(z^{2}) ,
\end{equation}
where ${\mathcal {P}}$ denotes the Cauchy principal value,
we obtain
\begin{equation}
\label{eq:GDiepsilon1}
G_{\subD}^{+}(\rho=\rho'=0;\Delta t)=-\frac{1}{4\pi^{2}L^{2}}\frac{1}{\sin^{2}\Delta t}-\frac{\rmi}{4\pi L^{2}}\delta(\sin^{2}\Delta t).
\end{equation}
Hence we can express the anti-commutator $G^{(1)}_{\subD}(x,x')=\langle 0|\left\{\hat{\varphi}(x),\hat{\varphi}(x')\right\}|0\rangle_{\subD}$ as
\begin{equation}
    G^{(1)}_{\subD}(\rho=\rho'=0;\Delta t)=-\frac{1}{4\pi^{2}L^{2}\sin^{2}\Delta t}.
\end{equation}
Note that this definition of the anti-commutator differs by a factor of two from the definition often employed. 
Now the Feynman Green's function can be expressed as
\begin{equation}
G_{\subD}(x,x')=\overline{G}(x,x')+\rmi G^{(1)}_{\subD}(x,x'),
\end{equation}
where $\overline{G}(x,x')$ is the average of the advanced and retarded Green functions. Note also that $\overline{G}(x,x')$ has support only on the lightcone so if we assume that $x$ and $x'$ are not connected by a null geodesic (as we have already assumed by separating only in the temporal direction) then we can ignore this term. The only contribution to the vacuum polarization comes from $G^{(1)}(x,x')$.

The Hadamard representation of the Feynman Green's function for a massless, conformally coupled scalar field on  CAdS (or AdS) takes the simple form \cite{Allen:1986ty,Kent:2014nya}
\begin{equation}
G_{\subS}(x,x')=\frac{\rmi}{4\pi^{2}}\frac{\Delta^{1/2}(x,x')}{(2\sigma+\rmi \epsilon)} ,
\label{eq:Hadamard}
\end{equation}
as the tail part vanishes for conformal fields. 
Here $\Delta^{1/2}(x,x')$ is the Van Vleck-Morette determinant which encodes information about the spray of neighbouring geodesics. 
For CAdS space-time, the Van Vleck-Morette determinant is a functional only of $\sigma$ which is known exactly in closed form \cite{Kent:2014nya}. 
We can also ignore the $\rmi \epsilon$ since this contributes only on the lightcone (for separated points) and we are assuming a temporal separation. 
For time-like separation, we have 
\begin{equation}
-2\sigma=L^{2}(\cos^{-1}Z)^{2}, \qquad Z=\frac{\cos\Delta t-\sin^{2}\rho}{\cos^{2}\rho} . 
\end{equation}
For $\rho=0$, this simply reduces to $-2\sigma=L^{2}\Delta t^{2}$ assuming small positive $\Delta t$. 
Similarly, the Van Vleck-Morette determinant is $\Delta^{1/2}=\Delta t^{3/2}\csc^{3/2}\Delta t$ for small positive $\Delta t$. 
Putting this together gives
\begin{equation}
G_{\subS}(\rho=\rho'=0;\Delta t)=-\frac{\rmi}{4\pi^{2}L^{2}}\left(\frac{1}{\Delta t^{2}}+\frac{1}{4}\right)+\Or(\Delta t^{2}).
\end{equation}
Similarly, the globally valid Feynman propagator expanded for small $\Delta t$ is
\begin{equation}
G_{\subD}(\rho=\rho'=0;\Delta t)=\rmi\,G^{(1)}_{\subD}(\rho=\rho'=0;\Delta t)=-\frac{\rmi}{4\pi^{2}L^{2}}\left(\frac{1}{\Delta t^{2}}+\frac{1}{3}\right)+\Or(\Delta t^{2}).
\end{equation}
Subtracting these and adopting the definition of the vacuum polarization gives
\begin{equation}
\langle 0|\hat{\varphi}^{2}|0\rangle_{\subD}=-\rmi\lim_{x'\to x}\left[G_{\subD}(x,x')-G_{\subS}(x,x')\right]=-\frac{1}{48\pi^{2}L^{2}}.
\end{equation}
This is precisely the answer one gets from the known closed-form representation which uses the maximal symmetry from the outset \cite{Allen:1986ty}. 
This calculation validates our mode-sum representation of the propagator.

An identical calculation gives for the Feynman Green's function with  Neumann boundary conditions
\begin{equation}
G_{\subN}(\rho=\rho'=0;\Delta t)=-\frac{\rmi}{4\pi^{2}L^{2}}\frac{\cos|\Delta  t|}{\sin^{2}(|\Delta t|-\rmi\epsilon)}=-\frac{\rmi}{4\pi^{2}L^{2}}\left(\frac{1}{\Delta t^{2}}-\frac{1}{6}\right)+\Or(\Delta t^{2}),
\end{equation}
where again we have ignored the Delta distribution piece that contributes only on the lightcone. Subtracting the local Hadamard representation as before gives
\begin{equation}
\langle 0|\hat{\varphi}^{2}|0\rangle_{\subN}=-\rmi\lim_{x'\to x}\left[G_{\subN}(x,x')-G_{\subS}(x,x')\right]=\frac{5}{48\pi^{2}L^{2}}.
\label{eq:vacpolN}
\end{equation}
Again this is precisely what one gets by assuming the propagator only depends on $\sigma$ from the outset \cite{Allen:1986ty}.

\subsection{Vacuum polarization for Robin boundary conditions}
\label{sec:VPRobinL}

We turn now to the mode-sum calculation of the vacuum polarization for the field satisfying arbitrary Robin boundary conditions. 
Since it is impossible to express the Feynman propagator in closed form  for general $\alpha $, one must compute the vacuum polarization by regularizing mode-by-mode. 
In other words, rather than express the Feynman Green's function in closed form and subtract the local Hadamard parametrix, we express the local Hadamard parametrix as a mode-sum and subtract from the Feynman Green's function mode-by-mode. 
While there are several recently-developed methods for achieving this in principle (see, for example, \cite{Freitas:2018mlu,Levi:2015eea,Levi:2016esr,Taylor:2016edd,Taylor:2017sux}), these methods are difficult to implement in the present situation since the mode-sum representation of the Hadamard parametrix is insensitive to the field boundary conditions and therefore the frequencies of such a decomposition are not those coming from the quantization condition (\ref{eq:RobinQuantization}). 

An alternative approach was employed in \cite{Barroso:2019cwp}, where Robin boundary conditions were applied to the $\ell=0$ modes only, all other field modes satisfying Dirichlet boundary conditions.
In \cite{Barroso:2019cwp}, the need to subtract the Hadamard parametrix in order to compute renormalized expectation values was circumvented by considering instead differences in expectation values between vacuum states for which the $\ell =0$ modes satisfy Robin boundary conditions, and all modes (including the $\ell =0$ modes) satisfy Dirichlet boundary conditions.
Such differences do not require renormalization since the Hadamard parametrix (\ref{eq:Hadamard}) is independent of the quantum state under consideration.
However, the fact that the frequencies appearing in the mode-sum decomposition (\ref{eq:Wightman1}) for Robin boundary conditions are not the same as those for Dirichlet boundary conditions introduces considerable challenges in the numerical computation in \cite{Barroso:2019cwp}.

A further complication that arises from the fact that we do not have a closed-form representation of the propagator for general Robin boundary conditions is that, implicit in the expression (\ref{eq:Wightman1}) is an `$\rmi\epsilon$' prescription which encodes both the nonuniqueness of the Green function on Lorentzian space-time and is also needed to define the propagator as a distribution. Implementing this prescription to get the correct propagator with the correct short-distance behaviour is straightforward when we have a closed-form representation, for example, equations (\ref{eq:GDiepsilon}--\ref{eq:GDiepsilon1}) show how we implement this prescription for the field satisfying Dirichlet boundary conditions.  However, when we do not have a closed-form representation, implementing this prescription is tricky since the propagator contains singularities not regulated by the Hadamard parametrix. 
Indeed, the propagator is singular even when the points are separated (more precisely, the contributions coming from null geodesics connecting the two points diverge). In the mode-sum representation of the propagator, this is manifest as the nonconvergence of the modes even when the points are separated, whereby there are undamped oscillations contributing to the mode-sum at large frequency coming from pairs of points connected by null geodesics. A numerical prescription, called the ``self-cancelation'' generalised integral, is developed in \cite{Levi:2015eea} to cure this divergence in black hole spacetimes, which is tantamount to implementing an `$\rmi\epsilon$' prescription. Things are likely more difficult in the AdS case since the propagation of null geodesics is rather complicated by the nature of the boundary \cite{Gannot:2018jkg,Akhmedov:2018lkp}.

To circumvent these issues, in the next section we therefore adopt a different methodology, by considering the Euclidean section of CAdS rather than the Lorentzian space-time we have studied thus far. The Euclidean Green function is unique and automatically a well-defined distribution without the need for an `$\rmi\epsilon$' prescription. Moreover, since our space-time is static, there is a unique correspondence between the Euclidean Green function on the Euclidean section and the Feynman Green function on the Lorentzian spacetime.

\section{Quantum states on the Euclidean section}
\label{sec:Euclidean}

Since the computation of renormalized vacuum expectation values on CadS with Robin boundary conditions applied to the scalar field has proven to be very challenging from a practical point of view \cite{Barroso:2019cwp}, in this section we study thermal and vacuum states on the Euclidean section. 
Transforming to the Euclidean section has proved to be a powerful tool for the computation of renormalized expectation values on black hole space-times (see, for example, \cite{Taylor:2016edd,Taylor:2017sux,Candelas:1984pg,Howard:1984qp,Howard:1985yg,Anderson:1994hg,Breen:2018ukd,Morley:2018lwn}), and
we will see that this greatly simplifies our computations.
In particular, we will be able to apply Robin boundary conditions to all field modes and compute the renormalized vacuum polarization for both vacuum and thermal states.

\subsection{The Euclidean Green's function}
\label{eq:EucGreen}

We perform the standard Wick rotation $\tau = -\rmi t$ and consider a quantum scalar field on the Euclidean space-time
\begin{equation}
    \rmd s^{2}=L^{2}\sec^{2}\rho\,\left( \rmd \tau^{2}+\rmd \rho^{2}+\sin^{2}\rho \, \rmd\Omega_{2}^{2}\right).
    \label{eq:metricE}
\end{equation}
Vacuum and thermal expectation values can then be computed as follows:
\begin{subequations}
\label{eq:EuclideanVP}
\begin{eqnarray}
\langle 0|\hat{\varphi}^{2}|0\rangle_{\alpha}
& = & 
\lim_{x'\to x}\left[G_{\alpha}^{\subE}(x,x')-G^{\subE}_{\subS}(x,x')\right] ,
\\
\langle \beta|\hat{\varphi}^{2}|\beta\rangle_{\alpha}
& = & 
\lim_{x'\to x}\left[G_{\alpha,\beta}^{\subE}(x,x')-G^{\subE}_{\subS}(x,x')\right] ,
\end{eqnarray}
\end{subequations}
where the superscript $\mathrm{E}$ refers to quantities constructed on the Euclidean space-time (\ref{eq:metricE}).
As previously, $|0\rangle_{\alpha}$ denotes a vacuum state with the scalar field satisfying Robin boundary conditions, while we use the notation $|\beta\rangle_{\alpha}$ to denote a thermal state at inverse temperature $\beta $, again with Robin boundary conditions applied.

For a thermal state at temperature $T$, the time coordinate $\tau$ is assumed to be periodic with periodicity $2\pi\beta=2\pi/T$. 
The temperature $T$ here is arbitrary, in other words, there exist thermal states satisfying the Hadamard condition at any temperature. This is in contrast to the Euclidean version of a black hole space-time, where there is a natural temperature associated with the black hole horizon and $2\pi T$ is the surface gravity of the black hole. 
Returning to CAdS, unlike the Lorentzian calculation, the periodicity in Euclidean ``time'' forces a discrete integer frequency spectrum independent of the boundary conditions imposed on the field. 
Hence the thermal Euclidean Green's function assumes the mode-sum representation
\begin{equation}
 \fl   G^{\subE}_{\alpha,\beta }(x,x')=\frac{\kappa }{8\pi ^{2} L^{2}}\cos\rho\,\cos\rho'
 \sum_{n=-\infty}^{\infty}\rme^{\rmi  n\kappa \Delta\tau }\sum_{\ell=0}^{\infty}(2\ell+1)P_{\ell}(\cos\gamma)g_{\omega \ell}(\rho,\rho')
 \label{eq:betaGE}
\end{equation}
where $\omega = n\kappa $ is the quantized frequency, $\gamma $ is the angular separation of the space-time points (\ref{eq:cosgamma}), $g_{\omega \ell}(\rho,\rho')$ is the one-dimensional Green's function satisfying the inhomogeneous equation
\begin{equation}
 \fl  \left\{\frac{\rmd}{\rmd\rho}\left(\sin^{2}\rho\frac{\rmd}{\rmd\rho}\right)-\omega ^{2} \sin^{2}\rho-\ell(\ell+1)\right\}g_{\omega \ell}(\rho,\rho')=\delta(\rho-\rho'), 
 \label{eq:oneDgeqn}
\end{equation}
and we have introduced the quantity
\begin{equation}
    \kappa = 2\pi T
    \label{eq:defkappa}
\end{equation}
to make the notation a little more compact.
For vacuum states, the coordinate $\tau $ is not periodic and the frequency is not quantized.
Hence the vacuum Euclidean Green's function has the mode-sum representation
\begin{equation}
 \fl  \hspace{0.5cm} G^{\subE}_{\alpha }(x,x')=\frac{1}{8\pi ^{2} L^{2}}\cos\rho\,\cos\rho'
 \int _{\omega =-\infty}^{\infty}\rmd \omega \, \rme^{\rmi  \omega \Delta\tau }\sum_{\ell=0}^{\infty}(2\ell+1)P_{\ell}(\cos\gamma)g_{\omega \ell}(\rho,\rho') .
 \label{eq:vacuumGE}
\end{equation}

The one-dimensional Green's function $g_{\omega \ell}(\rho,\rho')$ is constructed from a normalized product of solutions of the homogeneous version of (\ref{eq:oneDgeqn}),
\begin{equation}
\label{eq:normproduct}
    g_{\omega \ell}(\rho,\rho')=\frac{p_{\omega \ell}(\rho_{<})q_{\omega \ell}(\rho_{>})}{N_{\omega \ell}^{\rm {E}}}, 
\end{equation}
where $p_{\omega \ell}(\rho)$ is the solution which is regular  at the origin $\rho=0$,  the function $q_{\omega \ell}(\rho)$ is the solution satisfying the boundary conditions at the CAdS boundary $\rho=\pi/2$,
and $N_{\omega \ell}^{\rm {E}}$ is a normalization constant. 
We have adopted the notation $\rho_{<}\equiv\min\{\rho,\rho'\}$ and $\rho_{>}\equiv\max\{\rho,\rho'\}$. 
The general solution of the homogeneous version of (\ref{eq:oneDgeqn}) can be expressed in terms of  Conical (Mehler) functions as
\begin{equation}
    p_{\omega \ell }, q_{\omega \ell }\sim (\sin\rho)^{-1/2}\left[ C_{1}P^{-\ell-1/2}_{\rmi  \omega -1/2}(\cos\rho)+C_{2}P^{-\ell-1/2}_{\rmi  \omega -1/2}(-\cos\rho)\right] ,
\end{equation}
where $C_{1}$ and $C_{2}$ are arbitrary constants. 
Imposing regularity at the origin $\rho =0$ requires
\begin{equation}
    p_{\omega \ell}(\rho)=(\sin\rho)^{-1/2}P^{-\ell-1/2}_{\rmi  \omega -1/2}(\cos\rho),
\end{equation}
the overall constant being irrelevant since it can be absorbed into a redefinition of $N_{\omega \ell}^{\subE}$. 
It is at the boundary $\rho = \frac{\pi }{2}$ where there is freedom to choose boundary conditions. 
Taking, without loss of generality, 
\begin{equation}
q_{\omega \ell}= (\sin\rho)^{-1/2}\left[ C_{\omega \ell}^{\alpha}\,P^{-\ell-1/2}_{\rmi  \omega -1/2}(\cos\rho)+P^{-\ell-1/2}_{\rmi \omega -1/2}(-\cos\rho) \right] ,
\end{equation}
where $C_{\omega \ell}^{\alpha}$ is a constant, and imposing Robin boundary conditions on $q_{\omega \ell}$ analogous to (\ref{eq:Robin}),
\begin{equation}
    q_{\omega \ell}(\rho )+\frac{\rmd q_{\omega \ell}(\rho )}{\rmd\rho}  \tan\alpha =0 , \qquad \rho \rightarrow \pi/2,
    \label{eq:RobinE}
\end{equation}
fixes the constant $C_{\omega \ell}^{\alpha}$ to be
\begin{equation}
    C_{\omega \ell}^{\alpha}=\frac{2|\Gamma(\frac{\rmi  \omega +\ell+2}{2})|^{2}\tan\alpha-|\Gamma(\frac{\rmi  \omega +\ell+1}{2})|^{2}}{2 |\Gamma(\frac{\rmi  \omega +\ell+2}{2})|^{2}\tan\alpha+|\Gamma(\frac{\rmi  \omega +\ell+1}{2})|^{2}}.
    \label{eq:Comegaellalpha}
\end{equation}
This reduces to $C_{\omega \ell}^{\subD}=-1$ for Dirichlet boundary conditions whence
\begin{equation}
    q_{\omega \ell}^{\subD}(\rho)=(\sin\rho)^{-1/2}\left[P_{\rmi  \omega -1/2}^{-\ell-1/2}(-\cos\rho)-P_{\rmi  \omega -1/2}^{-\ell-1/2}(\cos\rho)\right] ,
\end{equation}
while $C_{\omega \ell}^{\subN}=1$ for Neumann boundary conditions and
\begin{equation}
    q_{\omega \ell}^{\subN}(\rho)=(\sin\rho)^{-1/2}\left[P_{\rmi  \omega -1/2}^{-\ell-1/2}(-\cos\rho)
    +P_{\rmi \omega -1/2}^{-\ell-1/2}(\cos\rho)\right] .
\end{equation}
It is useful to reexpress the function $q_{\omega \ell }^{\alpha }$ for general Robin boundary conditions in terms of a combination of these two special cases as
\begin{equation}
 q_{\omega \ell}^{\alpha}(\rho)
 =q_{\omega \ell}^{\subD}(\rho) \cos^{2}\alpha
 +q_{\omega \ell}^{\subN}(\rho) \sin^{2}\alpha
 +\left(C_{\omega \ell}^{\alpha}+\cos 2\alpha \right)
 (\sin\rho)^{-1/2}P_{\rmi  \omega -1/2}^{-\ell-1/2}(\cos\rho).
\end{equation}
We will see below that the benefit of this particular form is that all the divergences in the Euclidean Green's function come from the first two terms here; the mode-sum involving the last term is finite in the coincidence limit. 

The final step in the construction of the mode-sum representation of the Euclidean Green's function is computing the normalization constant in (\ref{eq:normproduct}). 
In order for $g_{\omega \ell}(\rho,\rho')$ to be a Green's function, we must have $N_{\omega \ell}^{\subE}=\sin^{2}\rho\,\mathcal{W}\{p_{\omega \ell},q_{\omega \ell}\}$ where $\mathcal{W}$ denotes the Wronskian of the solutions. 
This is straightforwardly calculated to be
\begin{equation}
    N_{\omega \ell}^{\subE}=\frac{2}{|\Gamma(\ell+1+\rmi \omega )|^{2}}.
\end{equation}
Note that this is independent of $\alpha$. 

Putting all of this together, and after some algebra, we obtain the following useful expressions for the Euclidean Green's functions for vacuum and thermal states:
\begin{subequations}
\label{eq:EuclideanGreenFn}
\begin{eqnarray}
\fl G_{\alpha}^{\subE}(x,x')& = &
G_{\subD}^{\subE}(x,x')\cos^{2}\alpha 
+G_{\subN  }^{\subE}(x,x')\sin^{2}\alpha
+G_{\subR }^{\subE}(x,x') \sin 2\alpha ,
\label{eq:EGFvac}
\\
\fl G_{\alpha,\beta }^{\subE}(x,x') & = &
G_{\subD,\beta }^{\subE}(x,x')\cos^{2}\alpha
+G_{\subN , \beta }^{\subE}(x,x')\sin^{2}\alpha
+G_{\subR ,\beta }^{\subE}(x,x') \sin 2\alpha  ,
\label{eq:EGFthermal}
\end{eqnarray}
\end{subequations}
where $G_{\subD}^{\subE}(x,x')$, $G_{\subD ,\beta }^{\subE}(x,x')$ are the Euclidean Green's functions for Dirichlet boundary conditions given by
\begin{subequations}
\label{eq:GDirichlet}
\begin{eqnarray}
\label{eq:GDvac}
 \fl   G_{\subD }^{\subE}(x,x')&=&
 \frac{1}{16\pi ^{2} L^{2}}\frac{\cos\rho\cos\rho'}{\sqrt{\sin\rho\,\sin\rho'}}\int _{\omega =-\infty}^{\infty}\rmd \omega \, \rme^{\rmi  \omega \Delta\tau}
 \sum_{\ell=0}^{\infty}(2\ell+1)P_{\ell}(\cos\gamma)
 |\Gamma(\ell+1+\rmi  \omega)|^{2}\nonumber\\
 \fl   & & \qquad  \times 
 P_{\rmi  \omega -1/2}^{-\ell-1/2}(\cos\rho_{<})
 \left[P_{\rmi \omega -1/2}^{-\ell-1/2}(-\cos\rho_{>})
 -P_{\rmi  \omega -1/2}^{-\ell-1/2}(\cos\rho_{>})\right],
\\
\label{eq:GDthermal}
 \fl   G_{\subD ,\beta }^{\subE}(x,x') & = &
 \frac{\kappa }{16\pi ^{2} L^{2}}\frac{\cos\rho\cos\rho'}{\sqrt{\sin\rho\,\sin\rho'}}
 \sum_{n=-\infty}^{\infty}\rme^{\rmi n \kappa\Delta\tau}\sum_{\ell=0}^{\infty}(2\ell+1)P_{\ell}(\cos\gamma)|\Gamma(\ell+1+\rmi n \kappa)|^{2}\nonumber\\
 \fl   & & \qquad \times P_{\rmi n\kappa -1/2}^{-\ell-1/2}(\cos\rho_{<})\left[P_{\rmi n \kappa-1/2}^{-\ell-1/2}(-\cos\rho_{>})-P_{\rmi n \kappa-1/2}^{-\ell-1/2}(\cos\rho_{>})\right],
\end{eqnarray}
\end{subequations}
$G_{\subN}^{\subE}(x,x')$, $G_{\subN , \beta }^{\subE}(x,x')$ are the Euclidean Green's function for Neumann boundary conditions given by
\begin{subequations}
\label{eq:GNeumann}
\begin{eqnarray}
 \fl   
 G_{\subN}^{\subE}(x,x')
 & = 
 &\frac{1}{16\pi ^{2}L^{2}}
 \frac{\cos\rho\cos\rho'}{\sqrt{\sin\rho\,\sin\rho'}}
 \int_{\omega =-\infty}^{\infty}\rmd \omega \, 
 \rme^{\rmi  \omega\Delta\tau}
 \sum_{\ell=0}^{\infty}(2\ell+1)P_{\ell}(\cos\gamma)
 |\Gamma(\ell+1+\rmi  \omega )|^{2}
 \nonumber \\
 \fl  & & \qquad
 \times P_{\rmi  \omega -1/2}^{-\ell-1/2}(\cos\rho_{<})
 \left[P_{\rmi \omega -1/2}^{-\ell-1/2}(-\cos\rho_{>})
 +P_{\rmi  \omega -1/2}^{-\ell-1/2}(\cos\rho_{>})\right],
 \label{eq:GNvac}
 \\
  \fl   G_{\subN, \beta }^{\subE}(x,x') & = &
  \frac{\kappa }{16\pi ^{2} L^{2}}\frac{\cos\rho\cos\rho'}{\sqrt{\sin\rho\,\sin\rho'}}\sum_{n=-\infty}^{\infty}\rme^{\rmi n \kappa\Delta\tau}\sum_{\ell=0}^{\infty}(2\ell+1)P_{\ell}(\cos\gamma)|\Gamma(\ell+1+\rmi n \kappa)|^{2}
\nonumber \\
 \fl  & &  \qquad \times  P_{\rmi n \kappa-1/2}^{-\ell-1/2}(\cos\rho_{<})\left[P_{\rmi n \kappa-1/2}^{-\ell-1/2}(-\cos\rho_{>})+P_{\rmi n \kappa-1/2}^{-\ell-1/2}(\cos\rho_{>})\right],
 \label{eq:GNthermal}
\end{eqnarray}
\end{subequations}
and $G_{\subR}^{\subE}(x,x')$, $G_{\subR, \beta }^{\subE}(x,x')$ are two-point functions (not Green's functions) whose mode-sum representations are
\begin{subequations}
\label{eq:GR}
\begin{eqnarray}
\fl G_{\subR}^{\subE}(x,x') & = & \frac{1}{16\pi ^{2}L^{2}}\frac{\cos\rho\cos\rho'}{\sqrt{\sin\rho\,\sin\rho'}}
\int_{\omega =-\infty}^{\infty} \rmd \omega \, 
\rme^{\rmi  \omega \Delta\tau}\sum_{\ell=0}^{\infty}(2\ell+1)
P_{\ell}(\cos\gamma)|\Gamma(\ell+1+\rmi  \omega )|^{2}
\nonumber\\ \fl & & 
\hspace{-1cm} \times  
\left[\frac{2|\Gamma(\frac{\rmi \omega +\ell+2}{2})|^{2}\cos\alpha
-|\Gamma(\frac{\rmi  \omega +\ell+1}{2})|^{2}\sin\alpha}{2 |\Gamma(\frac{\rmi  \omega +\ell+2}{2})|^{2}\sin\alpha
+|\Gamma(\frac{\rmi  \omega +\ell+1}{2})|^{2}\cos\alpha}\right]
P_{\rmi \omega -1/2}^{-\ell-1/2}(\cos\rho)
P_{\rmi \omega -1/2}^{-\ell-1/2}(\cos\rho'),
\nonumber \\
\fl \label{eq:GRvac}
\\
\fl G_{\subR,\beta }^{\subE}(x,x') & = & \frac{\kappa }{16\pi ^{2} L^{2}}\frac{\cos\rho\cos\rho'}{\sqrt{\sin\rho\,\sin\rho'}}
\sum_{n=-\infty}^{\infty}
\rme^{\rmi n \kappa\Delta\tau}\sum_{\ell=0}^{\infty}(2\ell+1)
P_{\ell}(\cos\gamma)|\Gamma(\ell+1+\rmi n \kappa)|^{2}
\nonumber\\
\fl & & \hspace{-1cm} \times \left[\frac{2|\Gamma(\frac{\rmi n \kappa+\ell+2}{2})|^{2}\cos\alpha-|\Gamma(\frac{\rmi n \kappa+\ell+1}{2})|^{2}\sin\alpha}{2|\Gamma(\frac{\rmi n \kappa+\ell+2}{2})|^{2}\sin\alpha +|\Gamma(\frac{\rmi n \kappa+\ell+1}{2})|^{2}\cos\alpha}\right]P_{\rmi n \kappa-1/2}^{-\ell-1/2}(\cos\rho)P_{\rmi n \kappa-1/2}^{-\ell-1/2}(\cos\rho').\nonumber\\\fl
\label{eq:GRthermal}
\end{eqnarray}
\end{subequations}

The two-point functions $G_{\subR}^{\subE}(x,x')$, $G_{\subR, \beta }^{\subE}(x,x')$ can be interpreted as the regular contributions to the
vacuum and thermal Green's function as a result of considering general Robin boundary conditions. 
These contributions are evidently vanishing for Dirichlet and Neumann boundary conditions by merit of the $\sin 2\alpha$ factor in (\ref{eq:EuclideanGreenFn}). 
In this sense, we can think of the subscript R as representing either `Robin' or `Regular'. 
To see that the mode-sums (\ref{eq:GR}) are indeed regular for any Green's functions (\ref{eq:EuclideanGreenFn}) satisfying the Hadamard condition, we note that both the Dirichlet (\ref{eq:GDirichlet}) and Neumann (\ref{eq:GNeumann}) Green's functions are known to satisfy the Hadamard condition. 
Hence the singularities in the first two terms of (\ref{eq:EuclideanGreenFn}) are given by $\cos^{2}\alpha\,G_{\subS}^{\subE}+\sin^{2}\alpha\,G_{\subS}^{\subE}=G_{\subS}^{\subE}$ where $G_{\subS}^{\subE}$ is the Hadamard parametrix for the Euclidean wave equation. 
Hence all the singularities for a propagator satisfying the Hadamard condition are contained in the first two terms of (\ref{eq:EuclideanGreenFn}), which implies that both $G_{\subR}^{\subE}$
and $G_{\subR,\beta }^{\subE}$ are regular in the coincidence limit.
This is in accordance with the fact that $G_{\subR}^{\subE}$
and $G_{\subR,\beta }^{\subE}$ are solutions of the homogeneous scalar field equation.

The contrapositive of the above argument is that if either $G_{\subR}^{\subE}$ or $G_{\subR,\beta }^{\subE}$ is not regular in the coincidence limit, then the corresponding Green's function $G_{\alpha}^{\subE}$ or $G_{\alpha,\beta }^{\subE}$ is not Hadamard.
In this case the corresponding quantum state is not a Hadamard state and should not be considered as physically meaningful. 
It is clear from the explicit mode-sum representation (\ref{eq:GR}) that $G_{\subR}^{\subE}$ or $G_{\subR,\beta }^{\subE}$ diverges if there exists a value of the constant $\alpha$ and mode numbers $(\omega ,\ell)$ for which
\begin{equation}
\label{eq:UnstableModes1}
    2\tan\alpha=
    -\frac{|\Gamma(\frac{\rmi \omega +\ell+1}{2})|^{2}}{|\Gamma(\frac{\rmi \omega +\ell+2}{2})|^{2}},
\end{equation}
which is precisely the condition (\ref{eq:UnstableModes}) for unstable modes.
In other words, the quantum state for the Robin boundary condition (\ref{eq:Robin}) is a Hadamard state only for those values of $\alpha $ for which the classical scalar field has no unstable modes.

\subsection{Equivalence of the Euclidean and Lorentzian Green's functions for Dirichlet and Neumann boundary conditions}
\label{sec:quasiclosed}

One advantage of the representation (\ref{eq:EuclideanGreenFn}) is that we will be able to use known expressions for the Dirichlet and Neumann propagators to simplify the Green's function for general Robin boundary conditions. 
The thermal propagator on CAdS for Dirichlet and Neumann boundary conditions can be obtained as an infinite image sum of the corresponding 
zero-temperature Green's function on the Lorentzian space-time \cite{Allen:1986ty}. 
It is not at all obvious how to connect that expression to the mode-sum representations (\ref{eq:GDirichlet}, \ref{eq:GNeumann}) derived here using Euclidean methods. 
In this section we therefore present the details of this calculation, proving that,
for Dirichlet and Neumann boundary conditions, the thermal Euclidean Green's functions  (\ref{eq:GDthermal}, \ref{eq:GNthermal}) are equivalent to the anticommutators for the field at finite temperature on the Lorentzian space-time under the mapping $\Delta t\to \rmi \Delta\tau$. 

We will show in detail how to obtain the thermal anticommutator derived in \cite{Allen:1986ty} for the Dirichlet case from our mode-sum (\ref{eq:GDthermal}). 
The calculation for the Neumann case is almost identical. 
We start with the generalized addition theorem for Gegenbauer functions
$C_{\lambda }^{\xi }$ \cite{Durand1976}
\begin{eqnarray}
   \fl & & 
    C_{\lambda}^{\xi }(x\,x'-z(1-x^{2})^{1/2}(1-x'^{2})^{1/2})
   \nonumber \\  \fl & &   \qquad  =  
    \frac{\Gamma(2\xi -1)}{|\Gamma(\xi )|^{2}}
    \sum_{\ell=0}^{\infty}\frac{(-1)^{\ell}4^{\ell}\Gamma(\lambda-\ell+1)
    \Gamma(\ell+\xi )^{2}}{\Gamma(\lambda+2\xi +\ell)}
   \nonumber\\ \fl & & \qquad \qquad  \times 
   (2\ell+2\xi -1)(1-x^{2})^{\ell/2}(1-x'^{2})^{\ell/2}
   C_{\lambda-\ell}^{\xi +\ell}(x_{<})
   C_{\lambda-\ell}^{\xi +\ell}(x_{>})C_{\ell}^{\xi -1/2}(z),
\end{eqnarray}
where $x_{<}\equiv\min\{x,x'\}$ and $x_{>}\equiv\max\{x,x'\}$.
This is valid for any complex $\lambda$ for which both sides of the equality are well-defined. Now taking $\xi =1$, $\lambda=\rmi n\kappa  -1$ and using the relationship between Legendre and Gegenbauer functions gives
\begin{eqnarray}
\fl 
\frac{1}{\sqrt{\sin\rho\sin\rho'}}
\sum_{\ell=0}^{\infty}(2\ell+1)P_{\ell}(\cos\gamma)
\left|\Gamma(\ell+1+\rmi n \kappa)\right|^{2}P_{\rmi n \kappa-1/2}^{-\ell-1/2}(\cos\rho_{<})P_{\rmi n \kappa-1/2}^{-\ell-1/2}(-\cos\rho_{>})\nonumber\\
=-\sqrt{2\pi}\frac{ n \kappa}{\sinh \pi n \kappa}\frac{P^{-1/2}_{\rmi n \kappa-1/2}(\cos\Psi)}{\sqrt{\sin\Psi}} ,
\label{eq:step1}
\end{eqnarray}
where
\begin{equation}
    \Psi=\cos^{-1}\left(-\cos\rho\cos\rho'-\cos\gamma\,\sin\rho\sin\rho'\right).
    \label{eq:Psi}
\end{equation}
The particular conical functions appearing on the right-hand-side of (\ref{eq:step1}) reduce to
\begin{equation}
    \frac{P^{-1/2}_{\lambda+1/2}(\cos z)}{\sqrt{\sin z}}=\sqrt{\frac{2}{\pi}}\frac{1}{(\lambda+1)}\frac{\sin(\lambda+1)z}{\sin z}.
\end{equation}
We thus arrive at the following summation formula
\begin{eqnarray}
    \fl \frac{1}{\sqrt{\sin\rho\sin\rho'}}\sum_{\ell=0}^{\infty}(2\ell+1)P_{\ell}(\cos\gamma)\left|\Gamma(\ell+1+\rmi n \kappa)\right|^{2}P_{\rmi n \kappa-1/2}^{-\ell-1/2}(\cos\rho_{<})P_{\rmi n \kappa-1/2}^{-\ell-1/2}(-\cos\rho_{>})\nonumber\\
    =\frac{2\,\sinh n \kappa \Psi}{\sinh\pi n \kappa\,\sin\Psi}.
\end{eqnarray}
A similar development yields
\begin{eqnarray}
\label{eq:additiontheorem2}
    \fl \frac{1}{\sqrt{\sin\rho\sin\rho'}}\sum_{\ell=0}^{\infty}(2\ell+1)P_{\ell}(\cos\gamma)\left|\Gamma(\ell+1+\rmi n \kappa)\right|^{2}P_{\rmi n \kappa-1/2}^{-\ell-1/2}(\cos\rho)P_{\rmi n \kappa-1/2}^{-\ell-1/2}(\cos\rho')\nonumber\\
    =\frac{2\,\sinh n \kappa \Psi^{*}}{\sinh\pi n \kappa\,\sin\Psi^{*}},
\end{eqnarray}
where
\begin{equation}
    \Psi^{*}=\pi+\cos^{-1}\left(-\cos\rho\,\cos\rho'+\cos\gamma\,\sin\rho\,\sin\rho'\right).
    \label{eq:Psi*}
\end{equation}
The function $\cos^{-1}$ is defined to be the inverse of $\cos$ restricted to $[0,\pi]$ so that $\Psi^{*}\in[\pi,2\pi]$. 
Importantly, $\Psi^{*}\ne 0$ which implies there are no singularities in the contribution coming from this sum. 
Employing these addition theorems in (\ref{eq:GDthermal}) gives
\begin{equation}
   G_{\subD,\beta }^{\subE}(x,x')=\frac{\kappa }{8\pi^{2}L^{2}}\cos\rho\,\cos\rho'\sum_{n=-\infty}^{\infty}\rme^{-\rmi n \kappa\Delta\tau}\left\{\frac{\sinh n \kappa \Psi}{\sinh n \kappa \pi\,\sin\Psi}-\frac{\sinh n \kappa \Psi^{*}}{\sinh n \kappa \pi\,\sin\Psi^{*}}\right\}.
   \label{eq:step2}
\end{equation}
Concentrating on the first sum in (\ref{eq:step2}), we can use the identity
\begin{equation}
    \frac{2}{\pi}\sum_{k=1}^{\infty}\frac{(-1)^{k-1}k\sin k z}{k^{2}+\omega^{2}}=\frac{\sinh \omega z}{\sinh\omega \pi},\qquad -\pi<z<\pi
\end{equation}
to express the sum as
\begin{equation}
    \sum_{n=-\infty}^{\infty}\rme^{-\rmi n \kappa \Delta \tau}\frac{\sinh n \kappa \Psi}{\sinh n\kappa  \pi}=\frac{\Psi}{\pi}+\frac{4}{\pi}\sum_{k=1}^{\infty}(-1)^{k-1}k\sin k \Psi\sum_{n=1}^{\infty}\frac{\cos n \kappa\Delta\tau}{k^{2}+n^{2}\kappa^{2}},
\end{equation}
where we have swapped the order of summation. 
The $n$-modes can now be summed, resulting in
\begin{equation}
 \sum_{n=-\infty}^{\infty}\rme^{-\rmi n \kappa \Delta \tau}\frac{\sinh n \kappa \Psi}{\sinh n \kappa \pi}=-\frac{\beta}{\pi}\sum_{k=1}^{\infty}\sin k (\Psi-\pi)\,\frac{\cosh k (\case{\beta}{2}-\Delta\tau)}{\sinh(k\,\beta/2)},
 \end{equation}
where we have used the fact that $\beta^{-1}=T=\kappa/(2\pi)$. Finally using the identity
\begin{equation}
    \frac{\cosh k (\case{\beta}{2}-\Delta\tau)}{\sinh(k\,\beta/2)}=\rme^{-k\Delta\tau}+\frac{2\cosh k \Delta\tau}{\rme^{k\beta}-1},
\end{equation}
and assuming without loss of generality that $\Delta\tau>0$, we can employ the standard series
\begin{equation}
    \sum_{k=1}^{\infty}\rme^{-k \Delta\tau}\sin k z=\frac{1}{2}\frac{\sin z}{\cosh\Delta\tau-\cos z},\quad \Delta\tau>0,
\end{equation}
to obtain
\begin{equation}
\label{eq:Psisum}
   \sum_{n=-\infty}^{\infty}\rme^{-\rmi n \kappa \Delta \tau}\frac{\sinh n \kappa \Psi}{\sinh n \kappa \pi}=\frac{\beta}{2\pi}\frac{\sin\Psi}{\cosh\Delta\tau+\cos\Psi}-\frac{2\beta}{\pi}\sum_{k=1}^{\infty}(-1)^{k}\sin k \Psi\,\frac{\cosh k\Delta\tau}{\rme^{k\beta}-1}.
\end{equation}
An identical calculation gives the analogous result for the second sum in (\ref{eq:step2}) with the replacement $\Psi^{*}\to \Psi$. 
Hence we obtain the following representation for the thermal Euclidean Green's function with Dirichlet boundary conditions
\begin{eqnarray}
\label{eq:GDclosed}
 G_{\subD,\beta }^{\subE}(x,x')&=&\frac{\cos\rho\,\cos\rho'}{8\pi^{2}L^{2}}\Bigg\{\frac{1}{\cosh\Delta\tau+\cos\Psi}-\frac{1}{\cosh\Delta\tau+\cos\Psi^{*}}\nonumber\\
& &-4\sum_{k=1}^{\infty}\frac{(-1)^{k}\cosh k\Delta\tau}{\rme^{k\beta}-1}\left(\frac{\sin k \Psi}{\sin\Psi}-\frac{\sin k \Psi^{*}}{\sin\Psi^{*}}\right)\Bigg\}.
\end{eqnarray}
This is equivalent (modulo a factor of two due to our conventions), under the transformation $\Delta \tau\to\rmi\Delta t$, to the thermal anti-commutator derived in \cite{Allen:1986ty} using the fact that the thermal propagator is periodic in imaginary time and hence can be obtained as an infinite image-sum of the Lorentzian zero-temperature propagator. 

The thermal Euclidean Green's function for Neumann boundary conditions can be derived analogously to the treatment above and is given by
\begin{eqnarray}
\label{eq:GNclosed}
 G_{\subN,\beta }^{\subE}(x,x')&=&\frac{\cos\rho\,\cos\rho'}{8\pi^{2}L^{2}}\Bigg\{\frac{1}{\cosh\Delta\tau+\cos\Psi}+\frac{1}{\cosh\Delta\tau+\cos\Psi^{*}}\nonumber\\
& &-4\sum_{k=1}^{\infty}\frac{(-1)^{k}\cosh k\Delta\tau}{\rme^{k\beta}-1}\left(\frac{\sin k \Psi}{\sin\Psi}+\frac{\sin k \Psi^{*}}{\sin\Psi^{*}}\right)\Bigg\}.
\end{eqnarray}
We have therefore proven that the thermal Euclidean Green's functions for Dirichlet and Neumann boundary conditions are identical to those for thermal states on Lorentzian CAdS space-time. 

We can derive closed-form expressions for the vacuum Euclidean Green's functions (\ref{eq:GDvac}, \ref{eq:GNvac}) by taking the zero-temperature limit of (\ref{eq:GDclosed}, \ref{eq:GNclosed}), which corresponds to the limit in which the inverse temperature $\beta \rightarrow \infty $.
The sums over $k$ in (\ref{eq:GDclosed}, \ref{eq:GNclosed}) are uniformly convergent for $\beta > \Delta \tau $, and therefore the limit and summation can be interchanged, giving
\begin{subequations}
\label{eq:Gvacclosed}
\begin{eqnarray}
 G_{\subD }^{\subE}(x,x')&=&\frac{\cos\rho\,\cos\rho'}{8\pi^{2}L^{2}}\Bigg\{\frac{1}{\cosh\Delta\tau+\cos\Psi}-\frac{1}{\cosh\Delta\tau+\cos\Psi^{*}}\Bigg\},
 \\
 G_{\subN }^{\subE}(x,x')&=&\frac{\cos\rho\,\cos\rho'}{8\pi^{2}L^{2}}\Bigg\{\frac{1}{\cosh\Delta\tau+\cos\Psi}+\frac{1}{\cosh\Delta\tau+\cos\Psi^{*}}\Bigg\}.
 \label{eq:GNeumannvac}
\end{eqnarray}
\end{subequations}

The advantage of expressing the thermal and vacuum Euclidean Green's functions for Dirichlet and Neumann boundary conditions in the quasi-closed forms (\ref{eq:GDclosed}, \ref{eq:GNclosed}, \ref{eq:Gvacclosed}) is that all of the singular terms, that is, the terms which contain all the short-distance divergences encoded in the Hadamard parametrix, are contained in the first term, which is in closed form. 
The remaining terms in each of the expressions (\ref{eq:GDclosed}, \ref{eq:GNclosed}, \ref{eq:Gvacclosed}) are finite in the coincidence limit.
This enables us to express the vacuum and thermal Euclidean Green's functions for general Robin boundary conditions in the same form, namely, by adopting (\ref{eq:GDclosed}, \ref{eq:GNclosed}, \ref{eq:Gvacclosed}) in (\ref{eq:EuclideanGreenFn}) we obtain
\begin{subequations}
\label{eq:GRclosed}
\begin{eqnarray}
\fl G_{\alpha}^{\subE}(x,x') & = & \frac{\cos\rho\,\cos\rho'}{8\pi^{2}L^{2}}\Bigg\{\frac{1}{\cosh\Delta\tau+\cos\Psi}-\frac{\cos2\alpha}{\cosh\Delta\tau+\cos\Psi^{*}}\Bigg\}
+G_{\subR}^{\subE}(x,x')\sin 2\alpha ,
\nonumber \\  \fl & & 
\\
\fl G_{\alpha ,\beta }^{\subE}(x,x')&=&\frac{\cos\rho\,\cos\rho'}{8\pi^{2}L^{2}}\Bigg\{\frac{1}{\cosh\Delta\tau+\cos\Psi}-\frac{\cos2\alpha}{\cosh\Delta\tau+\cos\Psi^{*}}\nonumber\\
\fl & &-4\sum_{k=1}^{\infty}\frac{(-1)^{k}\cosh k\Delta\tau}{\rme^{k\beta}-1}\left(\frac{\sin k \Psi}{\sin\Psi}-\cos2\alpha\,\frac{\sin k \Psi^{*}}{\sin\Psi^{*}}\right)\Bigg\}
\nonumber \\ \fl & & 
+G_{\subR, \beta }^{\subE}(x,x')\sin 2\alpha .
\end{eqnarray}
\end{subequations}

\subsection{Renormalized vacuum polarization}
\label{sec:VPE}

We turn now to the task of computing renormalized expectation values for the
vacuum polarization, which are given by (\ref{eq:EuclideanVP})
where 
$G_{\subS}^{\subE}(x,x')$ is the Hadamard parametrix (\ref{eq:Hadamard}). 
For the Euclideanized CAdS space-time, the Van Vleck-Morette determinant $\Delta^{1/2}(x,x')$ and Synge world function $\sigma(x,x')$ are given by
\begin{eqnarray}
    \Delta^{1/2}(x,x')&=&\frac{(2\sigma(x,x')/L^{2})^{3/4}}{\sinh^{3/2}(2\sigma(x,x')/L^{2})^{1/2}},\nonumber\\
    2\sigma(x,x')&=&L^{2}\left[\cos^{-1}\left(\frac{\cosh\Delta\tau-\cos\gamma\,\sin\rho\,\sin\rho'}{\cos\rho\,\cos\rho'}\right)\right]^{2}.
\end{eqnarray}
We are free to choose the direction in which we point-split so taking the spatial points together and splitting only in the temporal direction yields
\begin{equation}
    G_{\subS}^{\subE}(\Delta\tau;\rho)=\frac{1}{4\pi^{2}L^{2}}\frac{\cos^{2}\rho}{\Delta\tau^{2}}-\frac{1}{48\pi^{2}L^{2}}(2+\cos^{2}\rho)+\Or(\Delta\tau^{2}).
    \label{eq:HadamardE}
\end{equation}
To renormalize both the vacuum and thermal Euclidean propagators (\ref{eq:GRclosed}), it is sufficient to consider the first term since this contains all the singular parts.
Expanding this for temporal separation gives
\begin{equation}
\frac{\cos\rho\,\cos\rho'}{8\pi^{2}L^{2}}\frac{1}{\cosh\Delta\tau+\cos\Psi}
=
\frac{\cos^{2}\rho}{4\pi^{2}L^{2}}\left[ \frac{1}{\Delta\tau^{2}}-\frac{1}{12} \right] .
\label{eq:singbit}
\end{equation}
Subtracting (\ref{eq:HadamardE}) from (\ref{eq:singbit}) and taking the limit $\Delta\tau\to 0$ gives 
\begin{equation}
    \lim _{\Delta \tau \rightarrow 0} \left\{ 
    \frac{\cos^{2}\rho}{4\pi^{2}L^{2}}\left[ \frac{1}{\Delta\tau^{2}}-\frac{1}{12} \right]  - \frac{1}{4\pi^{2}L^{2}}\frac{\cos^{2}\rho}{\Delta\tau^{2}}+\frac{1}{48\pi^{2}L^{2}}(2+\cos^{2}\rho)
    \right\} 
    = \frac{1}{24\pi ^{2}L^{2}} .
\end{equation}
Therefore the vacuum and thermal expectation values for Robin boundary conditions are given by 
\begin{subequations}
\label{eq:phi2E}
\begin{eqnarray}
 \fl  \langle 0 | \hat{\varphi}^{2}(x)| 0\rangle_{\alpha} & = &
 -\frac{1}{48\pi^{2}L^{2}}(3\cos2\alpha-2) \nonumber\\
\fl & &  \hspace{-1cm}  +\frac{\sin2\alpha}{16\pi ^{2}  L^{2}}\frac{\cos^{2}\rho}{\sin\rho}
\int _{\omega =-\infty}^{\infty} \rmd \omega
\sum_{\ell=0}^{\infty}(2\ell+1)
|\Gamma(\ell+1+\rmi \omega )|^{2} D_{\omega \ell}^{\alpha}
\left[P_{\rmi  \omega -1/2}^{-\ell-1/2}(\cos\rho)\right]^{2},\nonumber\\
\fl & & 
\label{eq:phi2vac}
\\
 \fl  \langle \beta | \hat{\varphi}^{2}(x)|\beta\rangle_{\alpha}&=&-\frac{1}{48\pi^{2}L^{2}}(3\cos2\alpha-2)
 \nonumber \\ \fl & & 
 -\frac{\cos^{2}\rho}{2\pi^{2}L^{2}}\sum_{k=1}^{\infty}\frac{1}{\rme^{k\beta}-1}\left(k+(-1)^{k}\cos2\alpha\frac{\sin 2k \rho}{\sin 2\rho}\right)\nonumber\\
\fl & &  \hspace{-1cm}  +\frac{\sin2\alpha}{8\pi\beta L^{2}}\frac{\cos^{2}\rho}{\sin\rho}\sum_{n=-\infty}^{\infty}\sum_{\ell=0}^{\infty}(2\ell+1)|\Gamma(\ell+1+\rmi n \kappa)|^{2} D_{\omega \ell}^{\alpha}\left[P_{\rmi n \kappa-1/2}^{-\ell-1/2}(\cos\rho)\right]^{2}.\nonumber\\
\fl & & 
\label{eq:phi2thermal}
\end{eqnarray}
\end{subequations}
We remind the reader that, for thermal states, $\beta = T^{-1}=2\pi /\kappa $ and $\omega = n\kappa $. 
The constants $D_{\omega \ell}^{\alpha } $ are those in (\ref{eq:GR}):
\begin{equation}
    D_{\omega \ell }^{\alpha } = \frac{2|\Gamma(\frac{\rmi n \kappa+\ell+2}{2})|^{2}\cos\alpha-|\Gamma(\frac{\rmi n \kappa+\ell+1}{2})|^{2}\sin\alpha}{2|\Gamma(\frac{\rmi n \kappa+\ell+2}{2})|^{2}\sin\alpha +|\Gamma(\frac{\rmi n \kappa+\ell+1}{2})|^{2}\cos\alpha}.
    \label{eq:Domegaell}
\end{equation}
The expressions (\ref{eq:phi2E}) reduce to the expected results \cite{Allen:1986ty} when $\alpha =0$ and we have Dirichlet boundary conditions or $\alpha = \pi /2$ and Neumann boundary conditions are applied.
The results (\ref{eq:phi2E}) can be readily computed numerically. 
Away from the boundary $\rho = \pi /2$, all sums and integrals are rapidly convergent and the answers are dominated by the low-$\ell $, low-frequency modes. 

\begin{figure}
\centering
\begin{subfigure}{7.5cm}
\centering\includegraphics[width=7cm]{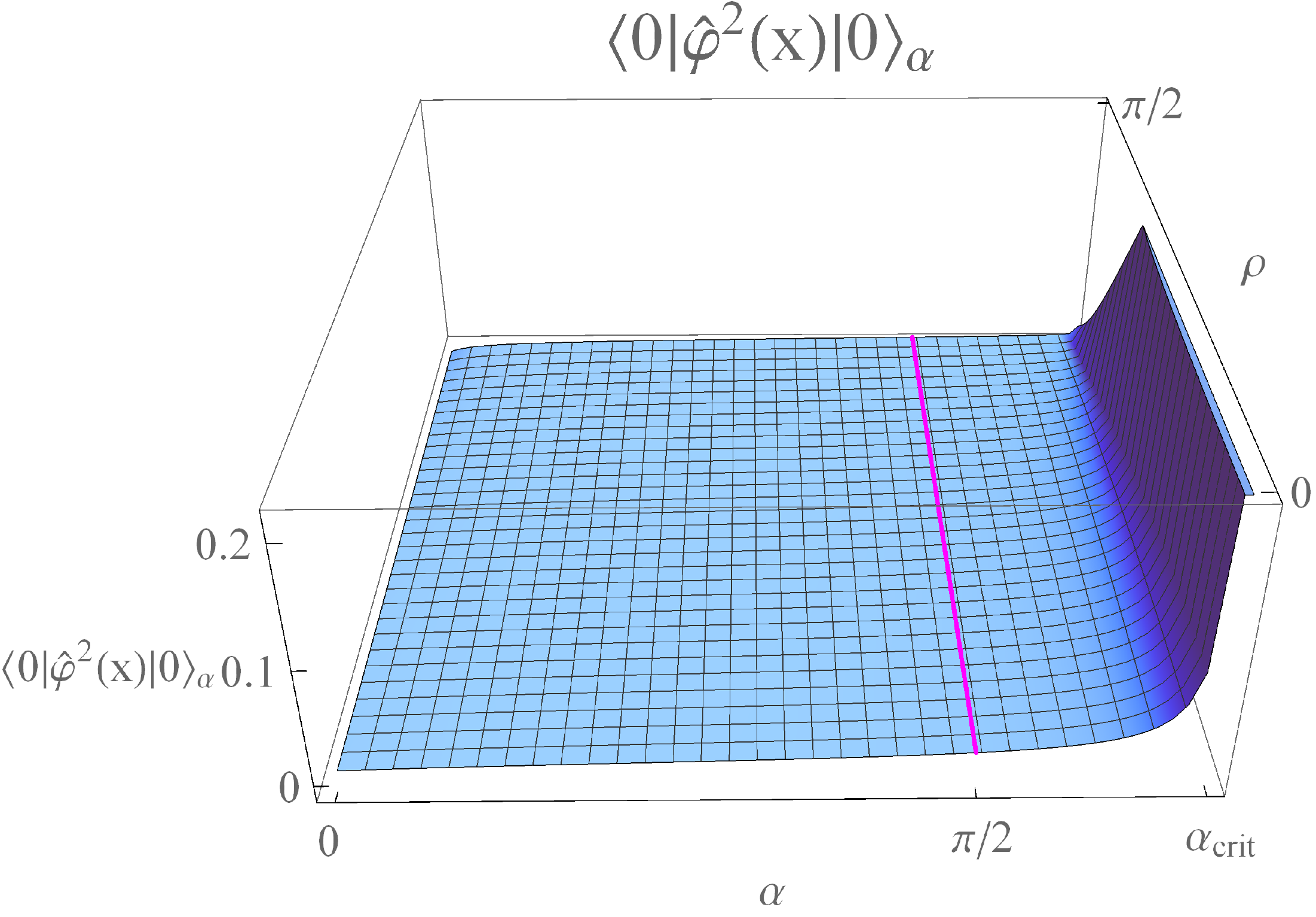}
\end{subfigure}%
\begin{subfigure}{7.5cm}
\centering\includegraphics[width=7cm]{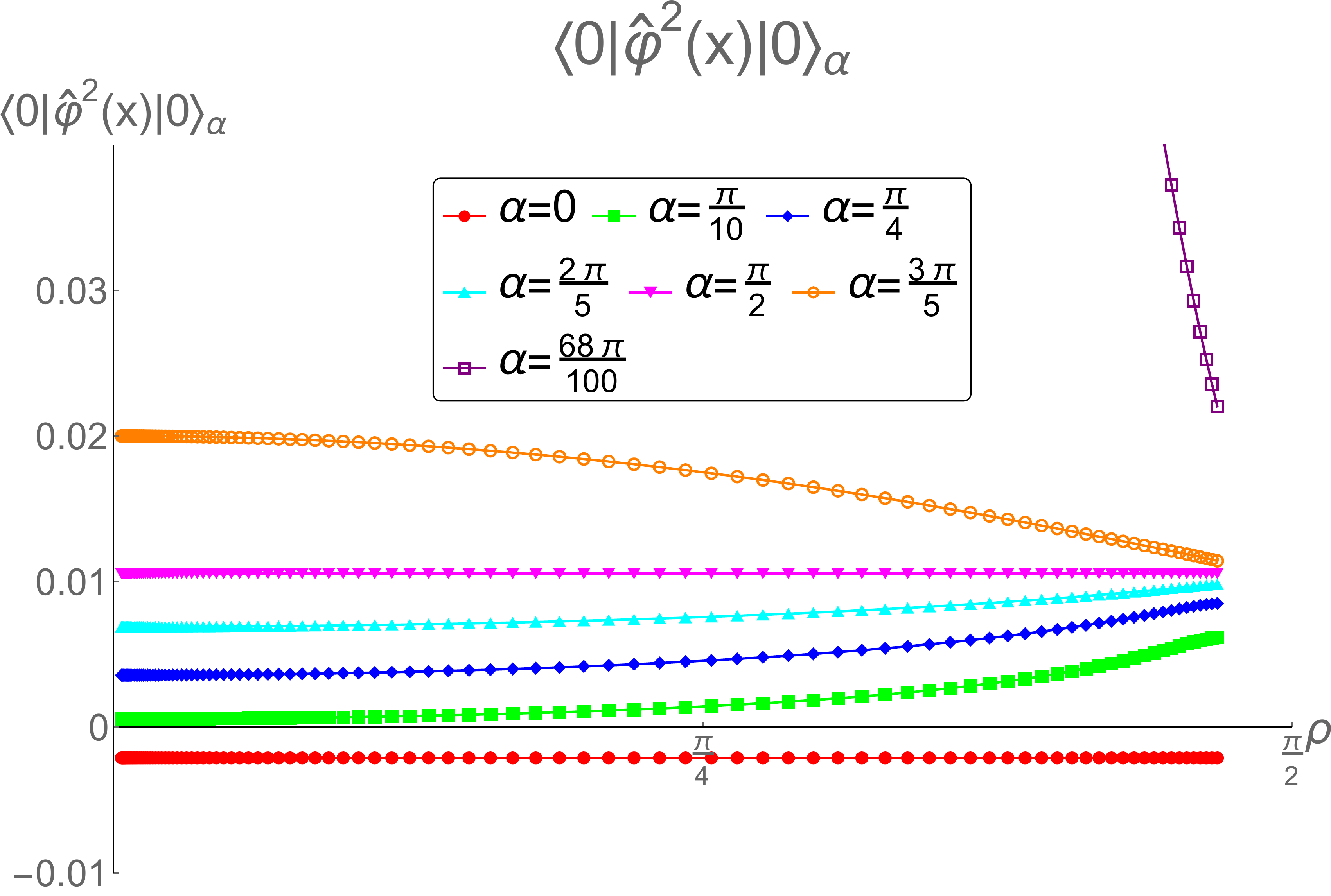}
\end{subfigure}
\caption{Vacuum expectation values   $\langle 0 | \hat{\varphi}^{2}(x)| 0\rangle_{\alpha}$ (\ref{eq:phi2vac}) as functions of the radial coordinate $\rho $ and the parameter $\alpha\in[0 ,\alpha_{\rm {crit}})$, where $\alpha_{{\rm {crit}}}$ is given by (\ref{eq:alphacrit}).
The left-hand figure shows a surface plot of $\langle 0 | \hat{\varphi}^{2}(x)| 0\rangle_{\alpha}$ as a function of $\rho $ and $\alpha $.
The magenta line marks the vacuum polarisation for Neumann boundary conditions (that is, $\alpha=\frac{\pi}{2}$), for which  $\langle 0 | \hat{\varphi}^{2}(x)| 0\rangle_{\pi /2}=\frac{5}{48\pi^{2}}$. 
The right-hand figure shows $\langle 0 | \hat{\varphi}^{2}(x)| 0\rangle_{\alpha}$ as a function of $\rho $  for some specific values of the parameter $\alpha$. We use units in which the AdS radius $L=1$.}
\label{fig:vac}
\end{figure}

We begin by studying the vacuum expectation values (\ref{eq:phi2vac}). 
In Figure~\ref{fig:vac}, the left-hand plot shows the vacuum expectation value as a function of the parameter $\alpha $ governing the Robin boundary conditions and the coordinate $\rho $.
In the right-hand plot, the profile of the vacuum expectation value as function of $\rho $ is shown for a selection of values of $\alpha $.
For Dirichlet ($\alpha =0$) and Neumann ($\alpha = \pi /2$) boundary conditions, the vacuum expectation value is a constant.
For all other values of $\alpha $, the vacuum expectation value is no longer constant as the boundary conditions have broken the maximal symmetry of the underlying CAdS space-time.
For $0<\alpha < \pi/2$, we find that the vacuum expectation value is monotonically increasing from the origin $\rho =0$ to the space-time boundary at $\rho =\pi /2$, while for $\pi /2<\alpha < \alpha _{\rm {crit}}$ the expectation values are monotonically decreasing away from the origin.
From the left-hand-plot, it is evident that
the values taken by $\langle 0 | \hat{\varphi}^{2}(x)| 0\rangle_{\alpha}$ at the origin $\rho =0$ increase monotonically with $\alpha $ as $\alpha $ increases. 
When $\alpha >\pi /2$ and approaches the critical value $\alpha _{\rm {crit}}$ (\ref{eq:alphacrit}), the vacuum expectation value increases rapidly at the origin.
This indicates the breakdown of the semiclassical approximation used here as $\alpha \rightarrow \alpha _{\rm {crit}}$, as anticipated due to the presence of classical instabilities when $\alpha _{\rm {crit}}<\alpha < \pi $.
The other striking feature, which can be clearly seen in the right-hand-plot, is that for all boundary conditions other than Dirichlet ($\alpha =0$), the vacuum expectation values approach the Neumann value $5/48\pi ^{2}L^{2}$ as $\rho \rightarrow \pi /2$ and the space-time boundary is approached.
We will discuss this further in section~\ref{sec:boundary}.

\begin{figure}
\centering
\begin{subfigure}{7.5cm}
\centering\includegraphics[width=7cm]{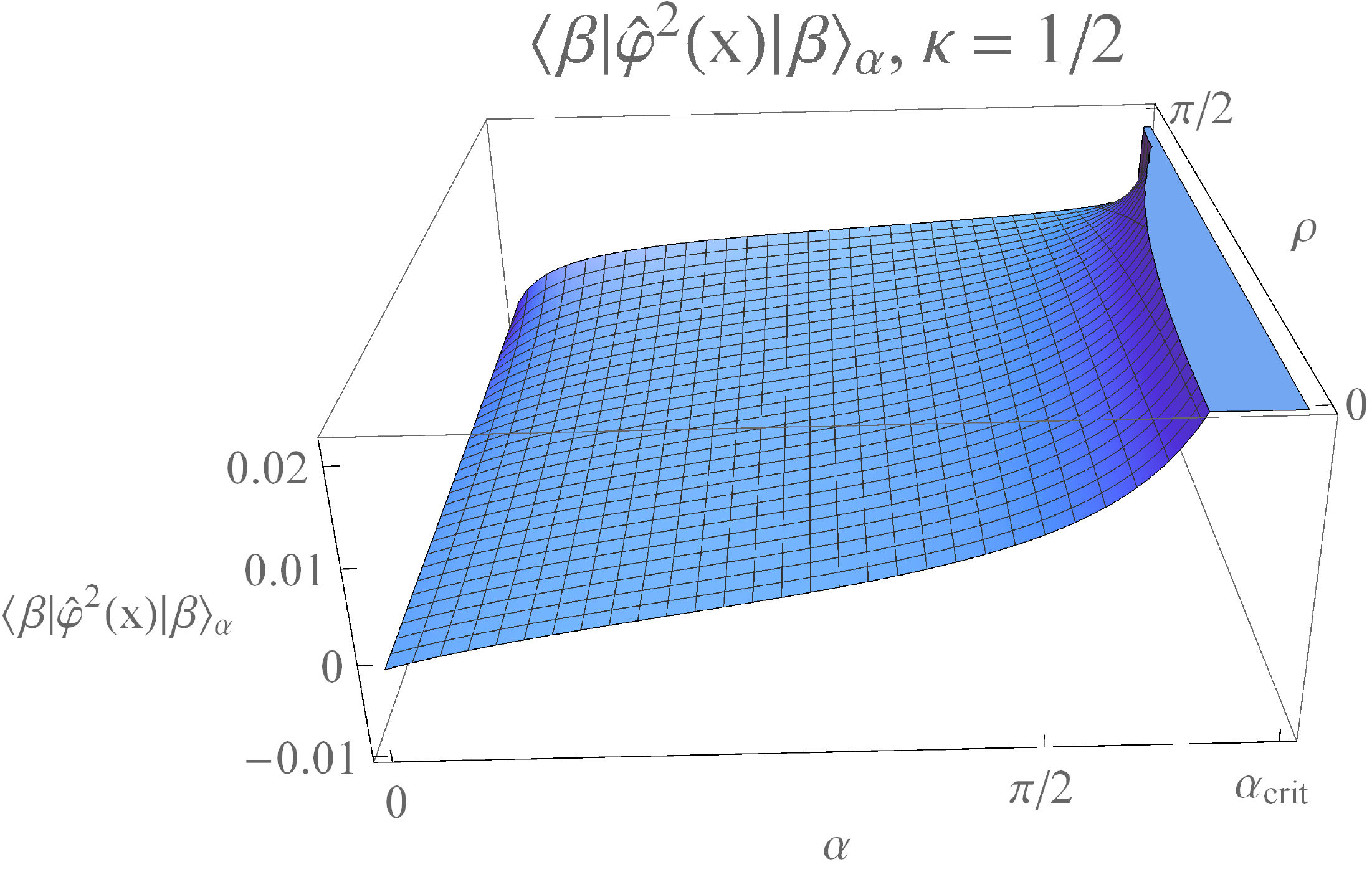}
\end{subfigure}%
\begin{subfigure}{7.5cm}
\centering\includegraphics[width=7cm]{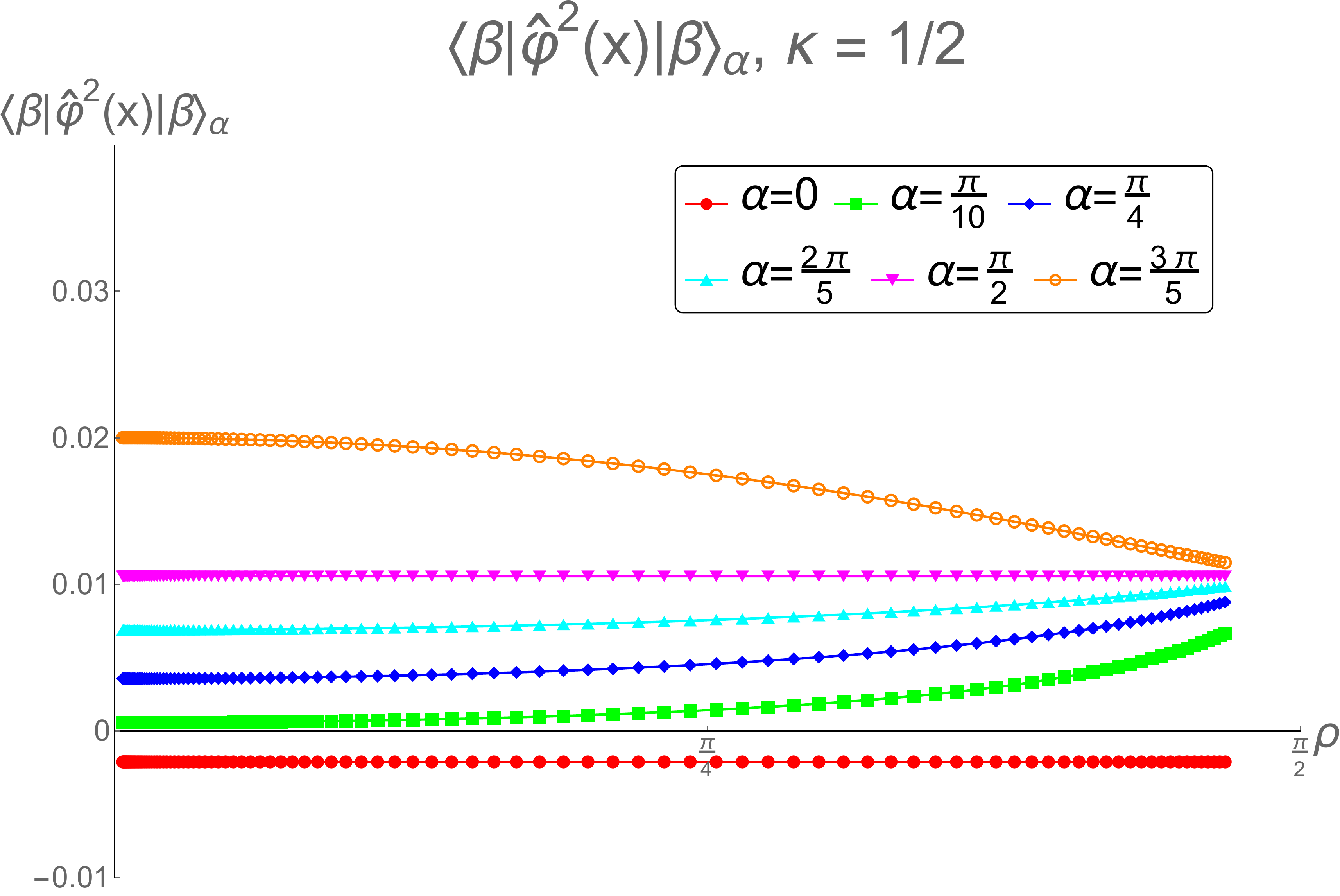}
\end{subfigure}\vspace{30pt}
\begin{subfigure}{7.5cm}
\centering\includegraphics[width=7cm]{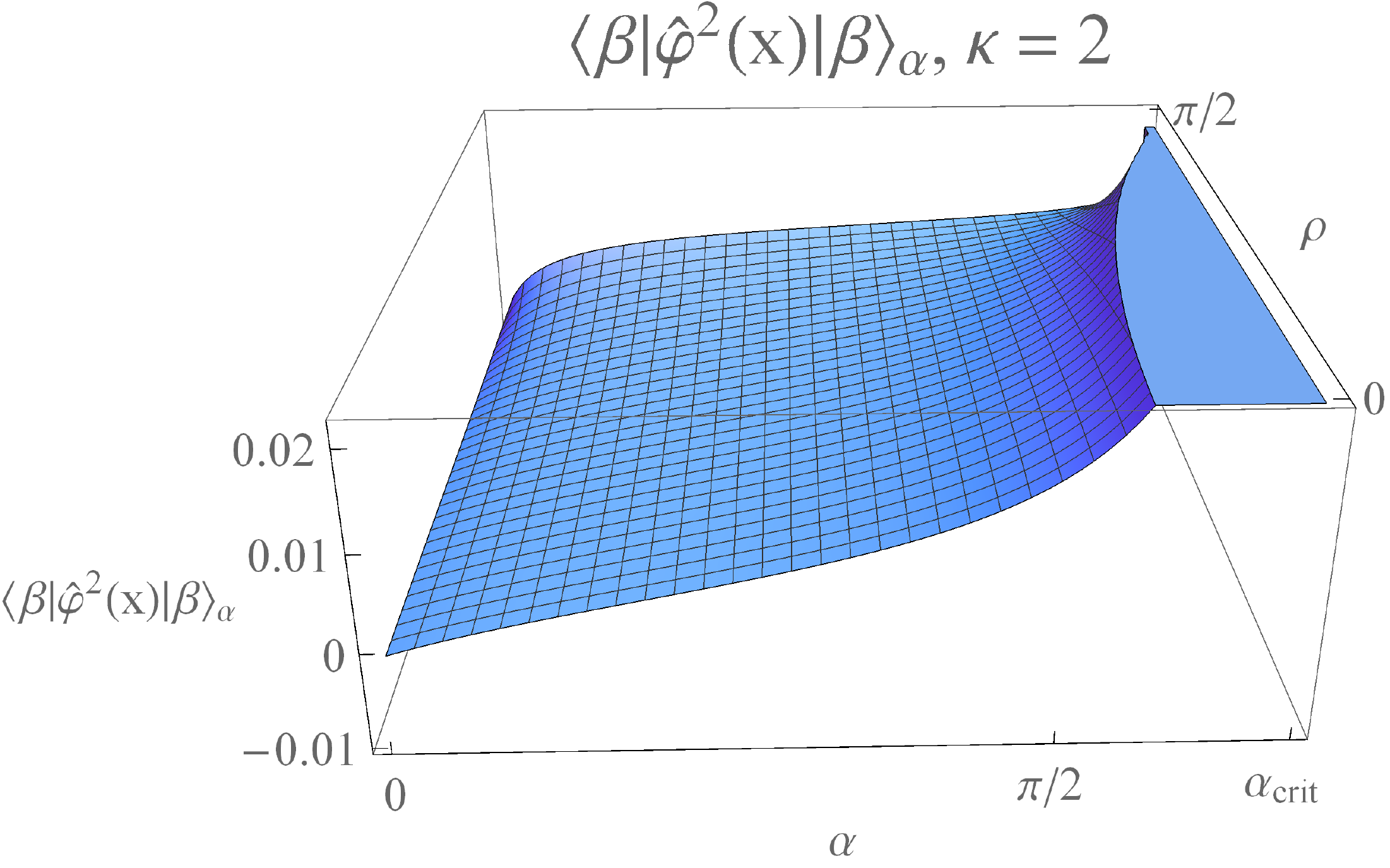}
\end{subfigure}%
\begin{subfigure}{7.5cm}
\centering\includegraphics[width=7cm]{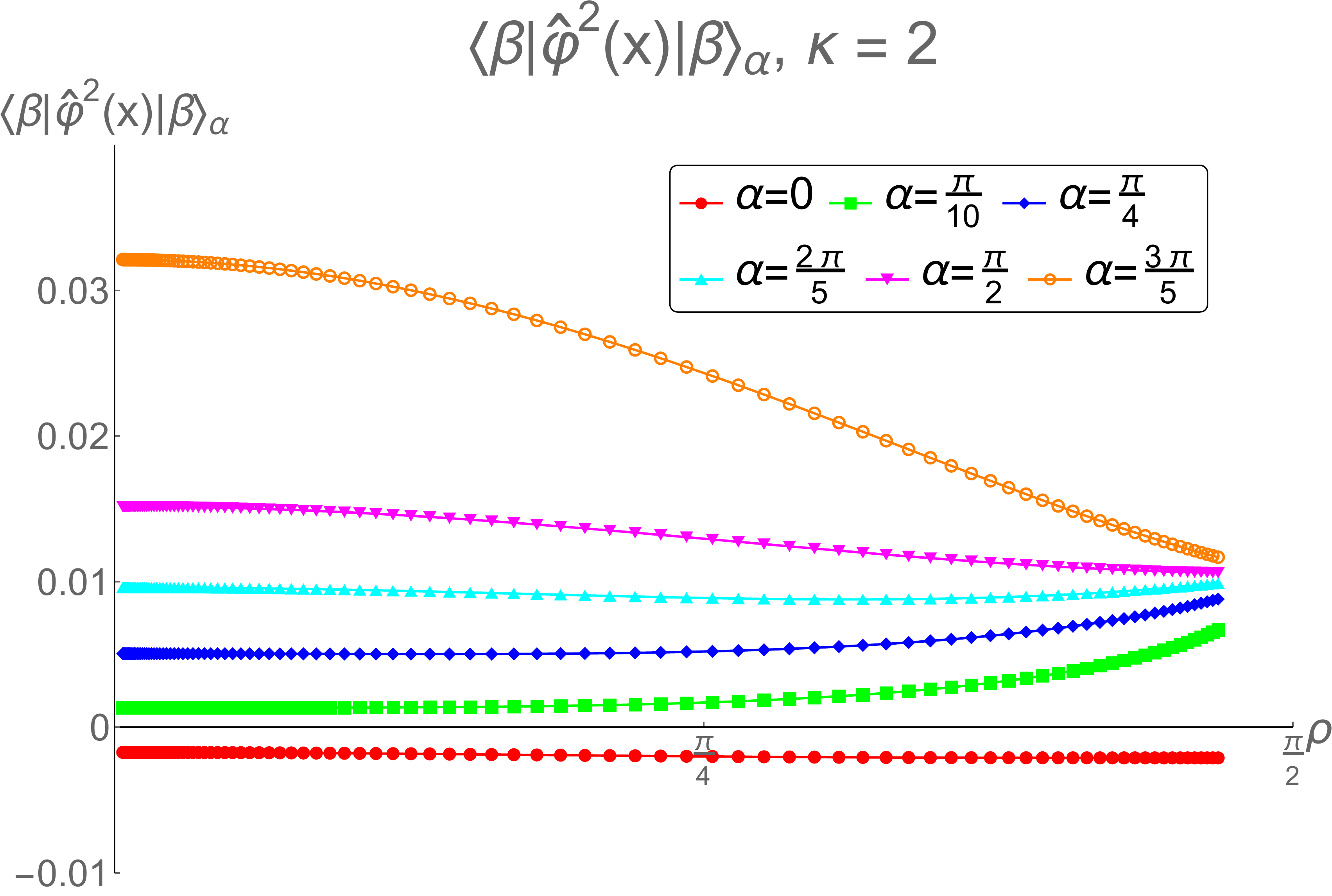}
\end{subfigure}\vspace{30pt}
\begin{subfigure}{7.5cm}
\centering\includegraphics[width=7cm]{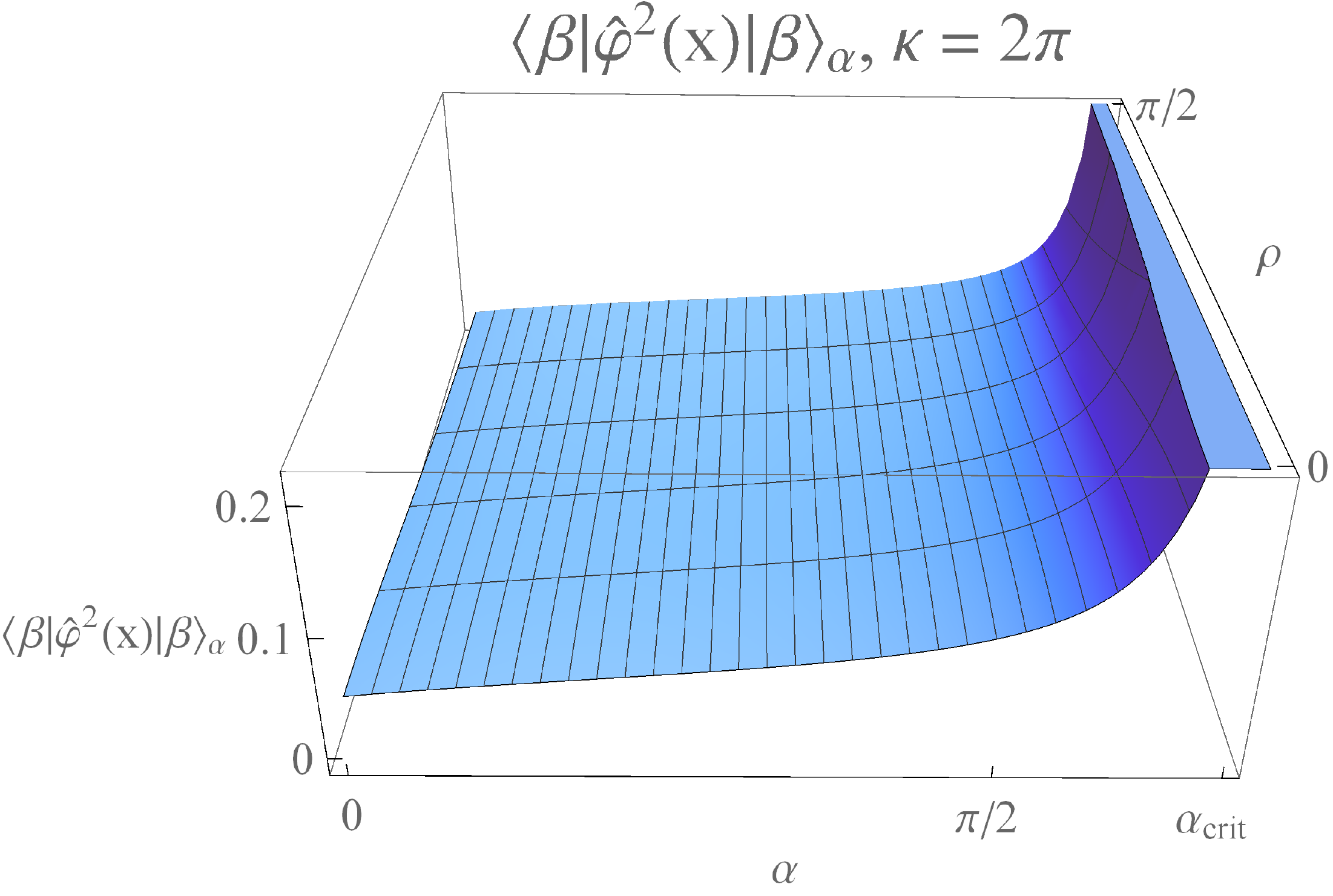}
\end{subfigure}%
\begin{subfigure}{7.5cm}
\centering\includegraphics[width=7cm]{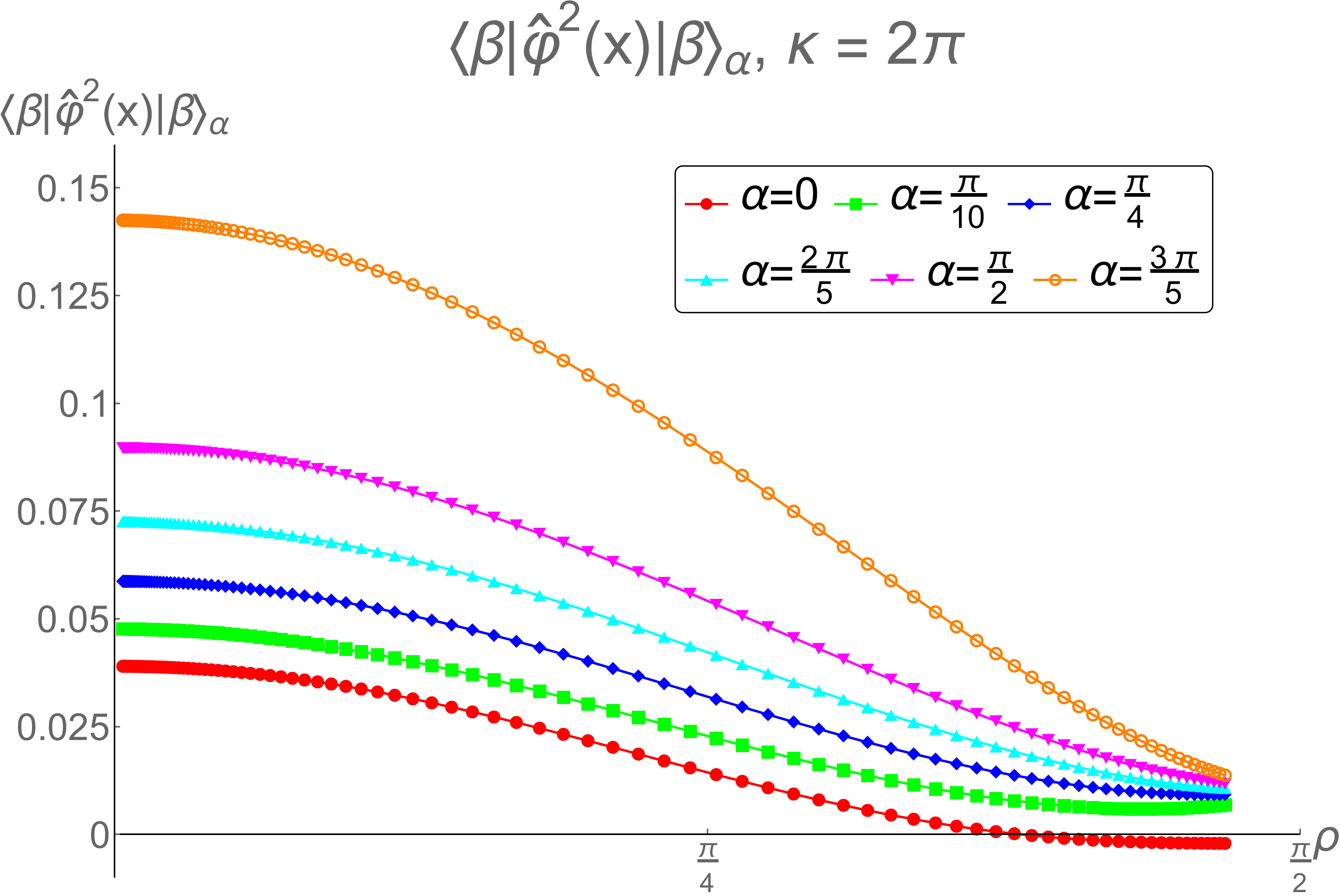}
\end{subfigure}
\caption{Thermal expectation values $\langle \beta | \hat{\varphi}^{2}(x)|\beta\rangle_{\alpha}$ (\ref{eq:phi2thermal}) 
as functions of the radial coordinate $\rho $ and the parameter $\alpha\in[0, \alpha_{\rm {crit}})$, where $\alpha_{{\rm {crit}}}$ is given by (\ref{eq:alphacrit}), for selected values of the temperature $T=\kappa /2 \pi $.
The left-hand figures show surface plots of $\langle \beta | \hat{\varphi}^{2}(x)| \beta \rangle_{\alpha}$ as a function of $\rho $ and $\alpha $.
The right-hand figures show $\langle \beta  | \hat{\varphi}^{2}(x)| \beta \rangle_{\alpha}$ as a function of $\rho $  for some specific values of the parameter $\alpha$. We use units in which the AdS radius $L=1$.}
\label{fig:thermal}
\end{figure}

We now turn to considering the thermal expectation values (\ref{eq:phi2thermal}), shown in Figure~\ref{fig:thermal}. 
The plots show the thermal expectation values for three selected values of the temperature $T=\kappa /2\pi $.
For each value of $T$, the left-hand plot shows the thermal expectation value as a function of the coordinate $\rho $ and the parameter $\alpha $, while the right-hand plot shows the profile as a function of $\rho $ for selected values of $\alpha $.

Consider first the thermal expectation values for small temperature, $\kappa = 1/2$.
In this case the profiles in the right-hand plot are virtually indistinguishable from those in Figure~\ref{fig:vac} for the vacuum expectation values.
The thermal expectation values for Dirichlet and Neumann boundary conditions are no longer constant in $\rho $, but the difference in values at the origin and infinity is extremely small and hence is not visible.
In the left-hand plot for $\kappa = 1/2$, we have used a different scale from that in Figure~\ref{fig:vac} for the vacuum expectation values.
The monotonically increasing behaviour of the thermal expectation values at the origin as $\alpha $ increases can be clearly seen.
When $\alpha =0$ and Dirichlet boundary conditions are applied, the thermal expectation values have their maximum at the origin and are monotonically decreasing as $\rho $ increases \cite{Allen:1986ty,Ambrus:2018olh}.
For $0<\alpha <\pi /2$, we find that the thermal expectation values are monotonically increasing as $\rho $ increases, while for $\pi /2\le \alpha < \alpha _{\rm {crit}}$ (including Neumann boundary conditions \cite{Allen:1986ty}) the thermal expectation values monotonically decrease as $\rho $ increases.

As the temperature increases, the thermal expectation value at the origin increases for all $\alpha $.
For all temperatures, at the origin the thermal expectation value is monotonically increasing as $\alpha $ increases and appears to diverge in the limit $\alpha \rightarrow \alpha _{\rm {crit}}$.
Away from the origin, the behaviour of the thermal expectation values is dependent upon both $\alpha $ and the temperature $T$.
For Dirichlet and Neumann boundary conditions, the thermal expectation value is monotonically decreasing as $\rho $ increases, for all values of the temperature \cite{Allen:1986ty}.
For sufficiently small temperatures and sufficiently small $\alpha >0$, we find that the thermal expectation value monotonically increases as $\rho $ increases from the origin to the space-time boundary. 
On the other hand, for sufficiently large $\alpha $, the thermal expectation value is monotonically decreasing as $\rho $ increases. 

In common with the vacuum expectation values, we see that the thermal expectation values approach the limit $5/48 \pi ^{2}L^{2}$ (the vacuum expectation value for Neumann boundary conditions) for all values of $\alpha $ except for $\alpha =0$, which corresponds to Dirichlet boundary conditions.
For all values of the temperature and parameter $\alpha $, we find that the thermal expectation values are larger than the vacuum expectation values for all values of the radial coordinate $\rho $, with this difference tending to zero as the space-time boundary is approached.
In all cases, the thermal radiation has ``clumped'' in a neighbourhood of the origin, due to the infinite gravitational potential at the space-time boundary.

\subsection{Vacuum polarization at the boundary}
\label{sec:boundary}

Our computation of the vacuum and thermal expectation values in the previous section has revealed an interesting feature.
Except for Dirichlet boundary conditions, as $\rho \rightarrow \pi /2$ and the space-time boundary is approached, all  expectation values seem to approach the limit $5/48 \pi ^{2}L^{2}$, which is the vacuum expectation value when Neumann boundary conditions are applied.
This behaviour is markedly different from that observed in \cite{Barroso:2019cwp} when Robin boundary conditions were applied only to the $\ell =0$ field modes. 
In that case all vacuum expectation values approached the Dirichlet value $-1/48\pi ^{2}L^{2}$ as $\rho \rightarrow \pi /2$.

Computing the sums in (\ref{eq:phi2E}) on the boundary $\rho=\pi/2$ turns out to be tricky since the sums are not all uniformly convergent.
The first sum (the $k$-sum) in the thermal expectation value (\ref{eq:phi2thermal}) converges for all $\rho$ and as a result of the overall $\cos^{2}\rho$ factor, this term vanishes on the CAdS boundary.
When $\alpha =0$ or $\pi /2$ and we are considering either Dirichlet or Neumann boundary conditions, the final sum in (\ref{eq:phi2thermal}) is absent and thermal expectation values coincide with vacuum expectation values on the boundary \cite{Allen:1986ty}.

\begin{figure}
    \centering
    \includegraphics[width=.8\linewidth]{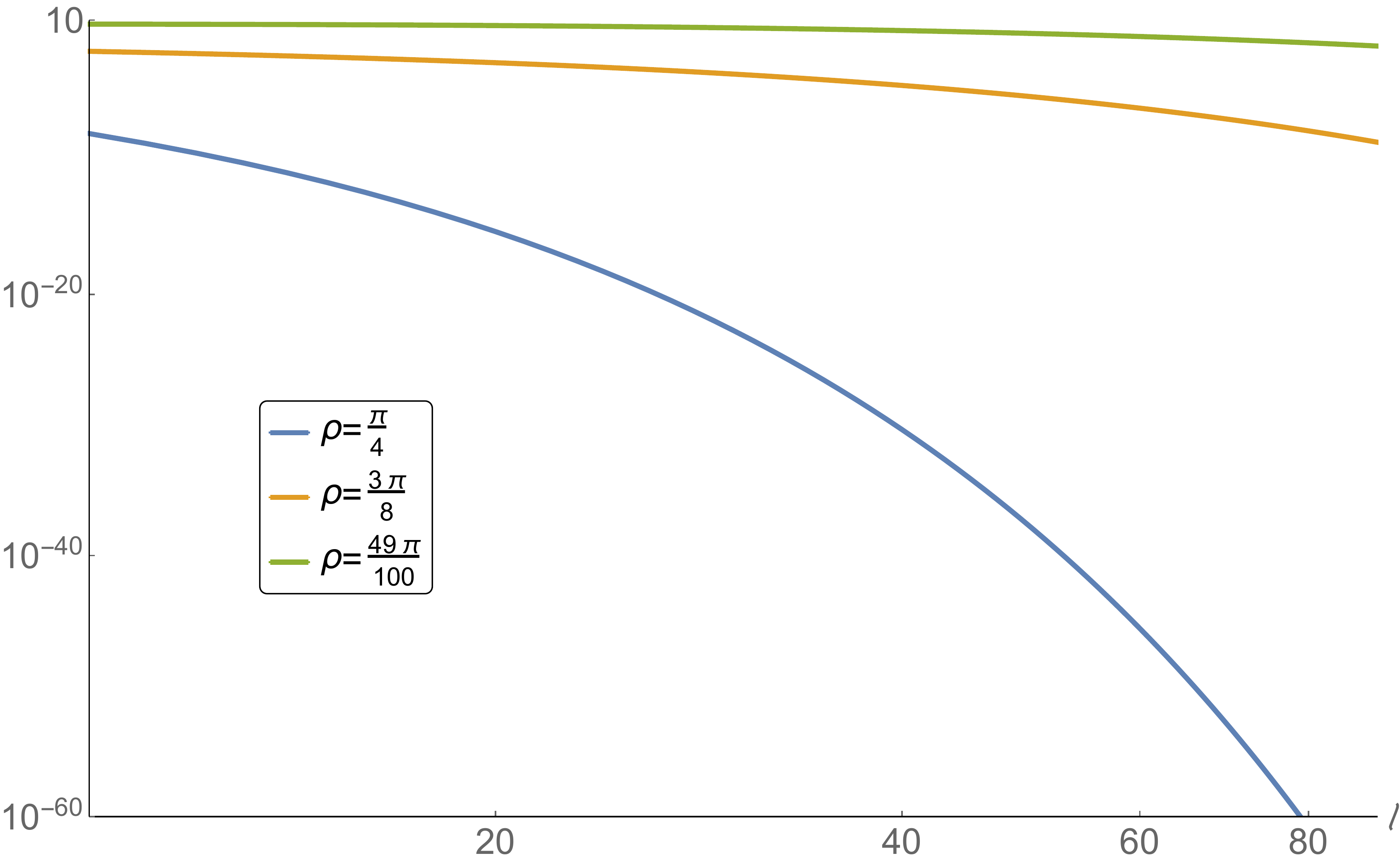}
    \caption{Log-log plot of the $\ell$-summand in the final sum in (\ref{eq:phi2thermal}), for a selection of values of the radial coordinate $\rho $. We have swapped the order of the sums, fixed $\alpha=\pi/4$, and performed the $n$-sum.}
    \label{fig:lconvergence}
\end{figure}

However, the last sum in (\ref{eq:phi2vac}, \ref{eq:phi2thermal}) is considerably more difficult to analyse for several reasons, including the fact that it is a double sum (or a sum and an integral in the vacuum case) and involves higher transcendental functions.
The sum over $\ell $ is not uniformly convergent in $\rho $, as can be seen in Figure~\ref{fig:lconvergence}.
For each fixed value of $0\le \rho <\pi /2$, the $\ell$-sum is convergent, but the rate of convergence decreases as $\rho $ increases towards the space-time boundary.
This means that we cannot naively interchange sums and limits to analyze the behaviour of the expectation values on the boundary.

In deriving (\ref{eq:phi2E}), we used the representations (\ref{eq:EuclideanGreenFn}) of the vacuum and thermal Green's functions with Robin boundary conditions applied in terms of the Green's functions with Dirichlet and Neumann boundary conditions, plus a correction term.
Our numerical results suggest that it will be useful, for all boundary conditions other than Dirichlet, to write (\ref{eq:EuclideanGreenFn}) as follows:
\begin{subequations}
\label{eq:EuclideanGreenFn1}
\begin{eqnarray}
\fl G_{\alpha}^{\subE}(x,x')& = &
G_{\subN  }^{\subE}(x,x')
+ \left[ G_{\subD  }^{\subE}(x,x') - G_{\subN  }^{\subE}(x,x') \right] \cos ^{2}\alpha 
+G_{\subR }^{\subE}(x,x') \sin 2\alpha ,
\label{eq:EGFvac1}
\\
\fl G_{\alpha,\beta }^{\subE}(x,x') & = &
G_{\subN , \beta }^{\subE}(x,x')
+ \left[ G_{\subD ,\beta  }^{\subE}(x,x') - G_{\subN ,\beta }^{\subE}(x,x') \right] \cos ^{2}\alpha 
+G_{\subR ,\beta }^{\subE}(x,x') \sin 2\alpha  .
\label{eq:EGFthermal1}
\end{eqnarray}
\end{subequations}
Using the expressions (\ref{eq:GDirichlet}, \ref{eq:GNeumann}, \ref{eq:GR}), we can write the vacuum and thermal expectation values with Robin boundary conditions applied in the alternative form
\begin{subequations}
\label{eq:phi2Ealt}
\begin{eqnarray}
 \fl  \langle 0 | \hat{\varphi}^{2}(x)| 0\rangle_{\alpha} & = &
\langle 0 | \hat{\varphi}^{2}(x)| 0\rangle_{\subN }
\nonumber\\
\fl & &  \hspace{-1cm} 
 -\frac{\sin2\alpha}{16\pi ^{2}  L^{2}}\frac{\cos^{2}\rho}{\sin\rho}
\int _{\omega =-\infty}^{\infty} \rmd \omega
\sum_{\ell=0}^{\infty}(2\ell+1)
|\Gamma(\ell+1+\rmi \omega )|^{2} E_{\omega \ell}^{\alpha}
\left[P_{\rmi  \omega -1/2}^{-\ell-1/2}(\cos\rho)\right]^{2}, \nonumber\\
\fl & & 
\label{eq:phi2vacalt}
\\
 \fl  \langle \beta | \hat{\varphi}^{2}(x)|\beta\rangle_{\alpha}&=&
 \langle \beta | \hat{\varphi}^{2}(x)|\beta\rangle_{\subN}
 \nonumber\\
\fl & &  \hspace{-1cm} 
 -\frac{\sin2\alpha}{8\pi\beta L^{2}}\frac{\cos^{2}\rho}{\sin\rho}\sum_{n=-\infty}^{\infty}\sum_{\ell=0}^{\infty}(2\ell+1)|\Gamma(\ell+1+\rmi n \kappa)|^{2} E_{\omega \ell}^{\alpha}\left[P_{\rmi n \kappa-1/2}^{-\ell-1/2}(\cos\rho)\right]^{2}, \nonumber\\
\fl & & 
\label{eq:phi2thermalalt}
\end{eqnarray}
\end{subequations}
where, in the thermal expectation values, the inverse temperature is $\beta = 2\pi /\kappa $ and $\omega = n\kappa $. 
We have defined new constants $E_{\omega \ell}^{\alpha}$ given by 
\begin{equation}
    E_{\omega \ell }^{\alpha } =  \cot \alpha - D_{\omega \ell }^{\alpha } 
    =
    \frac{\left| \Gamma \left( \frac{\rmi n\kappa + \ell + 1 }{2} \right)  \right| ^{2} \csc \alpha }{2\left| \Gamma \left( \frac{\rmi n\kappa + \ell + 2}{2} \right)  \right| ^{2} \sin \alpha + \left| \Gamma \left( \frac{\rmi n\kappa + \ell + 1 }{2} \right)  \right| ^{2} \cos \alpha } ,
    \label{eq:Eomegaell}
\end{equation}
where the constants $D_{\omega \ell }^{\alpha }$ can be found in (\ref{eq:Domegaell}).
The vacuum expectation value with Neumann boundary conditions applied can be found in (\ref{eq:vacpolN}) and the corresponding thermal expectation values are \cite{Allen:1986ty}
\begin{equation}
  \langle \beta | \hat{\varphi}^{2}(x)|\beta\rangle_{\subN} 
  = \langle 0 | \hat{\varphi}^{2}(x)| 0\rangle_{\subN }
-\frac{\cos^{2}\rho}{2\pi^{2}L^{2}}\sum_{k=1}^{\infty}\frac{1}{\rme^{k\beta}-1}\left(k-\frac{(-1)^k\sin 2k \rho}{\sin 2\rho}\right)  .
\end{equation}
As $\ell \rightarrow \infty $, we can derive the behaviour of the constants  $D_{n\ell }^{\alpha }$ (\ref{eq:Domegaell}) using the asymptotic properties of the $\Gamma $ functions \cite{NIST:DLMF}. 
We find that $D_{\omega \ell }^{\alpha }\rightarrow \cot \alpha $ as $\ell \rightarrow \infty $, providing that $\alpha > 0$.
This implies that the sums over  $\ell $ in (\ref{eq:phi2Ealt}) converge more rapidly than those in the last terms in (\ref{eq:phi2E}).
However, the sums in (\ref{eq:phi2Ealt}) are still not uniformly convergent, and so their behaviour cannot be easily deduced.

From our numerical investigations, it appears the final sum in each line of  (\ref{eq:phi2E}, \ref{eq:phi2Ealt}) diverges on the boundary. 
This is particularly delicate since we have a sum that appears to diverge and an overall $\cos^{2}\rho$ factor which vanishes on the boundary. 
The question then is whether the overall limit as the boundary is approached is finite and, if so, what the value of this limit might be.

Attempting to address this issue on CAdS is further complicated by the fact that the boundary $\rho =\pi /2$ is not part of the space-time.
We therefore consider instead the quantum scalar field on Euclideanized ESU, which has the additional advantage of being a globally hyperbolic space-time.
We consider the region of ESU for which $\rho \in [0,\pi /2]$, and impose Robin boundary conditions (\ref{eq:RobinE}) at $\rho =\pi /2$, which is now a surface in the space-time.
Since we are considering a massless, conformally coupled scalar field, the vacuum and thermal Euclidean Green's functions $G_{\alpha }^{\subESU}(x,x')$,
$G_{\alpha , \beta}^{\subESU}(x,x')$ on ESU are related to those on CAdS by 
\begin{equation}
G_{\alpha }^{\subE}(x,x') =  G_{\alpha }^{\subESU}(x,x')  \cos \rho \cos \rho ' , \qquad
    G_{\alpha , \beta }^{\subE}(x,x') =  G_{\alpha ,\beta }^{\subESU}(x,x')  \cos \rho \cos \rho '  .
    \label{eq:GESU}
\end{equation}
The problem of analyzing the behaviour of $G_{\alpha }^{\subE}(x,x')$ and
$G_{\alpha , \beta}^{\subE}(x,x')$ as the boundary is approached can then be tackled by studying the behaviour of $G_{\alpha }^{\subESU}(x,x')$ and $G_{\alpha ,\beta }^{\subESU}(x,x')$ as $\rho=\rho'\rightarrow \pi /2$ in ESU.

The divergences in the scalar Green's function which arise close to a boundary have been studied in \cite{Deutsch:1978sc}, and we now apply their general framework to our situation.
We begin by applying Stokes' Theorem (\ref{eq:stokes}) to the vacuum Euclidean Green's functions $G_{\subN }^{\subESU}(x,x')$ and $G_{\alpha }^{\subESU}(x,x')$ on the region $V \subset {\rm {ESU}}$ defined by $\rho \in [0, \pi/2] $:
\begin{eqnarray}
\fl    \int _{S} \left[ G_{\subN}^{\subESU}(x,y) {\widetilde {\nabla  }}_{\mu }G_{\alpha }^{\subESU}(y,x') - G_{\alpha}^{\subESU}(x,y) {\widetilde {\nabla  }}_{\mu }G_{\subN }^{\subESU}(y,x') \right] \, \rmd S^{\mu }
  \nonumber \\ 
  \fl & &  \hspace{-9cm} 
  = \int _{V} {\widetilde {\nabla  }}^{\mu } \left[ G_{\subN }^{\subESU}(x,y) {\widetilde {\nabla  }}_{\mu }G_{\alpha }^{\subESU}(y,x') - G_{\alpha}^{\subESU}(x,y) {\widetilde {\nabla  }}_{\mu }G_{\subN  }^{\subESU}(y,x') \right] \, \rmd V 
  \nonumber \\ 
  \fl & &  \hspace{-9cm} 
  = \int _{V} \left[ G_{\subN}^{\subESU}(x,y)  {\widetilde {\Box}}G_{\alpha }^{\subESU}(y,x') - G_{\alpha}^{\subESU}(x,y)  {\widetilde {\Box}}G_{\subN }^{\subESU}(y,x') \right] \, \rmd V , 
  \label{eq:StokesESU}
\end{eqnarray}
where $S=\partial V$ is the boundary of the region $V$, and all integrals are taken over the space-time points $y$.
The covariant derivatives ${\widetilde {\nabla }}$ and operator ${\widetilde {\Box }}={\widetilde {\nabla }}^{\mu } {\widetilde {\nabla }}_{\mu }$ are defined with respect to the Euclidean ESU metric
\begin{equation}
\rmd s^{2} = L^{2} \left[ \rmd \tau ^{2} + \rmd \rho ^{2} + \sin ^{2} \rho \, \rmd \Omega _{2}^{2} \right] .
\label{eq:ESUE}
\end{equation}
For all values of $\alpha $, the vacuum Euclidean Green's functions $G_{\alpha}^{\subESU}(x,y) $ satisfy the inhomogeneous version of the ESU scalar field equation (\ref{eq:ESUwave})
\begin{equation}
    \left\{ {\widetilde {\Box}} - \frac{1}{L^{2}} \right\} 
    G_{\alpha}^{\subESU}(x,x')  = - \frac{1}{{\sqrt {\widetilde {g}}}} \delta ^{(4)}(x,x'),
    \label{eq:inhomwave}
\end{equation}
where ${\widetilde {g}}$ is the determinant of the Euclidean ESU metric (\ref{eq:ESUE}), and $\delta ^{(4)}(x,x')$ is the four-dimensional Dirac delta function.
The vacuum Euclidean Green's functions  also satisfy the boundary conditions (\ref{eq:Robin}) at the space-time point $x$:
\begin{equation}
    G_{\alpha }^{\subESU}(x,x') \cos \alpha + \frac{\partial G_{\alpha }^{\subESU}(x,x')}{\partial \rho } \sin \alpha =0, \qquad \rho = \frac{\pi }{2}.
    \label{eq:ESUboundary}
\end{equation}
The boundary $S=\partial V$ of the region $V$ consists of two surfaces, $S={\widetilde {{\mathcal {I}}}}_{0} \cup {\widetilde {{\mathcal {I}}}}_{\pi/2}$, where ${\widetilde {{\mathcal {I}}}}_{0}$ is the time-like hypersurface at $\rho =0$ in ESU and ${\widetilde {{\mathcal {I}}}}_{\pi /2}$  at $\rho = \pi /2$. Both these hypersurfaces extend infinitely in the time direction, and are the ESU analogue of the hypersurfaces ${\mathcal {I}}_{0}$ and ${\mathcal {I}}_{\pi /2} $ in CAdS considered in section \ref{sec:canonical}.
It is shown in section \ref{sec:canonical} that each individual mode on CAdS does not contribute to the surface integral over ${\mathcal {I}}_{0}$.
Since the conformal factor $\cos \rho $ relating CAdS and ESU is equal to unity when $\rho =0$, each mode on ESU also does not contribute to the surface integral over ${\widetilde {\mathcal {I}}}_{0}$.
Writing the vacuum Euclidean Green's functions $G_{\alpha }^{\subESU}(x,x')$ as sums over modes, the contribution to (\ref{eq:StokesESU}) from the surface integral over ${\widetilde {\mathcal {I}}}_{0}$ is therefore zero.

The normal derivative to the boundary ${\widetilde {\mathcal {I}}}_{\pi /2}$ is simply $L^{-1}\partial /\partial \rho $, and therefore, applying the boundary conditions (\ref{eq:ESUboundary}) and using the inhomogeneous scalar field equation (\ref{eq:inhomwave}), the integrals in (\ref{eq:StokesESU}) become
\begin{equation}
\fl 
-L^{-1}\cot \alpha \int _{{\widetilde{\mathcal {I}}}_{\pi /2}} G_{\subN}^{\subESU}(x,y)G_{\alpha }^{\subESU}(y,x') \, \rmd S 
=
-G_{\subN}^{\subESU}(x,x')+G_{\alpha }^{\subESU}(x,x') ,
\end{equation}
where we have used the fact that $G_{\subN}^{\subESU}(x,x')$ satisfies the boundary condition (\ref{eq:ESUboundary}) with $\alpha = \pi /2$.
Rearranging, we find
\begin{equation}
    G_{\alpha }^{\subESU}(x,x') = G_{\subN}^{\subESU}(x,x')
    - L^{-1}\cot \alpha  \int _{{\widetilde{\mathcal {I}}}_{\pi /2}} G_{\subN}^{\subESU}(x,y)G_{\alpha }^{\subESU}(y,x') \, \rmd S .
    \label{eq:Deutsch1}
\end{equation}
Since the Neumann Green's function satisfies the Hadamard condition, for a general space-time point $x$ all the short-distance singularities in $G_{\alpha }^{\subESU}(x,x')$ are contained in the first term in (\ref{eq:Deutsch1}), and the integral will be finite in the limit $x'\rightarrow x$, except possibly if $x$ lies on the boundary.
The factor $\cot \alpha $ indicates that this expansion is valid for all Robin boundary conditions with $0<\alpha <\pi $ but not Dirichlet boundary conditions with $\alpha = 0$. 

Following \cite{Deutsch:1978sc}, the result (\ref{eq:Deutsch1}) forms the basis of an iterative expression for the vacuum Euclidean Green's function with Robin boundary conditions, by repeatedly inserting $G_{\alpha }^{\subESU}(x,x')$ into the integral on the right-hand-side of (\ref{eq:Deutsch1}).
This yields
\begin{equation}
   G_{\alpha }^{\subESU}(x,x') = G_{\subN}^{\subESU}(x,x')
   -L^{-1}  G_{\alpha }^{\subESU (1)}(x,x') \cot \alpha 
   +L^{-2} G_{\alpha }^{\subESU (2)}(x,x') \cot ^{2}\alpha  + \ldots ,
   \label{eq:ESUGalphaexp}
\end{equation}
where 
\begin{subequations}
\label{eq:GESUiterate}
\begin{eqnarray}
G_{\alpha }^{\subESU (1)}(x,x') & = & 
\int _{{\widetilde{\mathcal {I}}}_{\pi /2}} G_{\subN}^{\subESU}(x,y)G_{\subN}^{\subESU}(y,x') \, \rmd S , 
\label{eq:GESU1}
\\
G_{\alpha }^{\subESU (2)}(x,x') & = & 
\int _{{\widetilde{\mathcal {I}}}_{\pi /2}} G_{\subN}^{\subESU}(x,y)
\left[ 
\int _{{\widetilde {\mathcal {I}}}_{\pi /2}} G_{\subN}^{\subESU}(y,z)
G_{\subN}^{\subESU}(z,x') \, \rmd S \right]  \, \rmd S , 
\nonumber \\ & & 
\label{eq:ESUGalpha2}
\end{eqnarray}
\end{subequations}
where in (\ref{eq:ESUGalpha2}) the inner integral is over the space-time points $z$.
Subsequent terms in the expansion (\ref{eq:ESUGalphaexp}) contain additional integrals over ${\widetilde {\mathcal {I}}}_{\pi /2}$.

The Euclidean Green's function for Neumann boundary conditions, $G_{\subN}^{\subESU}(x,x')$ can be written in closed form using (\ref{eq:GNeumannvac}, \ref{eq:GESU})
\begin{equation}
G_{\subN }^{\subESU}(x,x')=\frac{1}{8\pi^{2}L^{2}}\Bigg\{\frac{1}{\cosh\Delta\tau+\cos\Psi}+\frac{1}{\cosh\Delta\tau+\cos\Psi^{*}}\Bigg\},
\label{eq:GNexact}
\end{equation}
where $\Psi $ and $\Psi ^{*}$ are given by (\ref{eq:Psi}, \ref{eq:Psi*}) respectively.
From (\ref{eq:GNexact}), it is straightforward to compute the renormalized vacuum polarization on ESU when Neumann boundary conditions are applied.
Considering points split only in the $\tau$-direction, we have $\cos \Psi =-1$, and $\cos \Psi ^{*}=\cos 2\rho $, from which $G_{\subN}^{\subESU}(x,x')$ is
\begin{eqnarray}
    G_{\subN}^{\subESU}(\Delta \tau, \rho, \theta, \varphi )
    & = & \frac{1}{8\pi ^{2}L^{2}} \left\{ 
    \frac{1}{\cosh \Delta \tau -1} + \frac{1}{\cosh \Delta \tau + \cos 2\rho }
    \right\}
    \nonumber \\
    & = & \frac{1}{8\pi ^{2}L^{2}} \left\{  \frac{2}{\Delta \tau ^{2}}
    + \left( \frac{1}{1+\cos 2\rho } - \frac{1}{6} \right) 
    \right\} + \Or(\Delta \tau ^{2}) .
\end{eqnarray}
To renormalize, we need to subtract the singular Hadamard terms $G^{\subESU}_{\subS}(x,x')$.
For our chosen point-splitting, Synge's world function on ESU is given by $2\sigma = L^{2}\Delta \tau ^{2}$ and the Van Vleck-Morette determinant on ESU is $\Delta ^{\frac{1}{2}}= 1 +\Or (\Delta \tau ^{3})$, so that
\begin{equation}
    G^{\subESU}_{\subS}(\Delta \tau , \rho ,\theta, \varphi ) =
    \frac{\Delta ^{\frac{1}{2}}}{8\pi ^{2}\sigma } 
    = \frac{1}{4\pi ^{2}L^{2}\Delta \tau ^{2}}+\Or (\Delta \tau ) .
    \label{eq:GHadESU}
\end{equation}
The renormalized vacuum expectation value on ESU for Neumann boundary conditions is then
\begin{eqnarray}
\langle 0 | \hat{\varphi}^{2}(x)| 0\rangle_{\subN }^{\subESU}
 & = & \lim _{\Delta \tau \rightarrow 0 }\left\{
G_{\subN}^{\subESU}(\Delta \tau, \rho, \theta, \varphi )
-G^{\subESU}_{\subS}(\Delta \tau , \rho ,\theta, \varphi )
 \right\}
 \nonumber \\ 
& = & \frac{5-\cos 2\rho }{48\pi ^{2}L^{2}\left( 1+ \cos 2\rho \right) }.
\label{eq:VPNESU}
\end{eqnarray}
Note that, unlike the corresponding renormalized vacuum expectation value on CAdS (\ref{eq:vacpolN}), this is not a constant on ESU. 
It should also be emphasised that, although the vacuum Euclidean Green's functions on CAdS and ESU are conformally related by (\ref{eq:GESU}), the relationship between Synge's world function $\sigma (x,x')$ and the Van Vleck-Morette determinant $\Delta ^{\frac{1}{2}}$ on CAdS and ESU is not so simple.
Therefore the Hadamard subtraction term $G_{\subS}^{\subESU}(x,x')$ is not conformally related to that on CAdS, as may be seen by comparing (\ref{eq:HadamardE}) and (\ref{eq:GHadESU}).

The vacuum expectation value  (\ref{eq:VPNESU}) is finite for all $0\le \rho <\pi /2$, but for $\rho = \pi/2-\epsilon$, it diverges in the limit $\epsilon \rightarrow 0$:
\begin{equation}
\langle 0 | \hat{\varphi}^{2}(x)| 0\rangle_{\subN }^{\subESU}
=     \frac{1}{16\pi ^{2}L^{2}\epsilon ^{2}}+ \Or (\epsilon ^{2}).
\end{equation}
This $\Or (\epsilon ^{-2})$ divergence as the boundary is approached is in accordance with the general analysis of \cite{Deutsch:1978sc}.
The work of \cite{Deutsch:1978sc} shows that the next term $G_{\alpha }^{\subESU (1)}(x,x')$ in the expansion (\ref{eq:ESUGalphaexp}) of the ESU Green's function is expected to diverge as $\Or (\epsilon ^{-1})$ as the boundary is approached, with subsequent terms being finite on the boundary.
Since we have a simple closed-form expression (\ref{eq:GNexact}) for $G_{\subN}^{\subESU}(x,x')$, we can test this general expectation by an explicit evaluation of the integral in (\ref{eq:GESU1}).

To perform the integral in (\ref{eq:GESU1}), we need to set one of the points on the boundary, so we fix $\rho '=\pi /2$ without loss of generality, in which case $\cos \Psi = \cos \Psi ^{*}=-\cos \gamma \sin \rho$, the two terms in  (\ref{eq:GNexact}) are equal and we have
\begin{equation}
   G_{\subN }^{\subESU}(\tau, \rho, \theta , \varphi ; \tau', \pi/2, \theta ' , \varphi ')=\frac{1}{4\pi^{2}L^{2}}\frac{1}{\cosh \Delta \tau -\cos \gamma \sin \rho  }. 
\end{equation}
Since $G_{\alpha }^{\subESU (1)}(x,x')$ is finite in the limit $x'\rightarrow x$, we set $x'=x$ in (\ref{eq:GESU1}) to give the integral
\begin{equation}
    G_{\alpha }^{\subESU (1)}(x,x) =
    \frac{1}{16\pi ^{4}L^{4}}
    \int _{{\widetilde {\mathcal {I}}}_{\pi /2}} \frac{1}{\left[ \cosh \Delta \tau - \cos \gamma \sin \rho \right] ^{2}} \, \rmd S.
\end{equation}
Here $x$ is a general point in ESU with coordinates $(\tau, \rho,  \theta, \phi )$, the integral is over space-time points $y=(\tau _{y},\pi /2, \theta _{y}, \varphi _{y})$ on the boundary, where $\Delta \tau = \tau _{y}-\tau $ and $\gamma $ is the angular separation of the points $x$ and $y$, given by (\ref{eq:cosgamma}) with $\theta '=\theta _{y}$ and $\Delta \varphi = \varphi _{y}-\varphi $.
Without loss of generality we may set $\theta =0$, $\varphi =0$ and then $\cos \gamma = \cos \theta _{y}$.
The integration over the angular variables can be performed to yield
\begin{equation}
   G_{\alpha }^{\subESU (1)}(x,x) 
   = \frac{1}{4\pi ^{3}L} \int _{\Delta \tau  = -\infty }^{\infty }
  \frac{1}{\cosh ^{2}\Delta \tau - \sin ^{2}\rho }\, \rmd \Delta \tau .
\end{equation}
The integrand is regular for all $\Delta \tau $ if $0\le \rho <\pi /2$, but singular at $\Delta \tau =0$ if $\rho = \pi /2$.
The integral over $\Delta \tau $ can be performed for $0<\rho<\pi/2 $, to give
\begin{equation}
 G_{\alpha }^{\subESU (1)}(x,x) = \frac{1}{2\pi ^{3}L} \frac{\rho }{\sin \rho \cos \rho }.
\end{equation}
For $\rho = \pi /2- \epsilon$, as $\epsilon \rightarrow 0 $ we therefore have 
\begin{equation}
    G_{\alpha }^{\subESU (1)}(x,x) = \frac{1}{4\pi ^{2}L\epsilon } + \Or (1) .
\end{equation}
As expected, this diverges like $\epsilon ^{-1}$ as the boundary is approached.
From \cite{Deutsch:1978sc}, the higher-order terms in the expansion (\ref{eq:ESUGalphaexp}) are all finite on the boundary so we do not need to consider them in detail.

We now apply this analysis to the renormalized expectation values on CAdS.
First, we multiply (\ref{eq:ESUGalphaexp}) by $\cos \rho \cos \rho '$ to give the following expression for the vacuum Euclidean Green's function $G_{\alpha }^{\subE}(x,x')$ (\ref{eq:GESU}) on CAdS with Robin boundary conditions applied:
\begin{equation}
   G_{\alpha }^{\subE}(x,x') = G_{\subN}^{\subE}(x,x')
    - L^{-1} \cos \rho \cos \rho ' \cot \alpha  \int _{{\widetilde{\mathcal {I}}}_{\pi /2}} G_{\subN}^{\subESU}(x,y)G_{\alpha }^{\subESU}(y,x') \, \rmd S .
\end{equation}
The second term is finite in the limit $x'\rightarrow x$, so we have the following expression for the renormalized vacuum expectation value:
\begin{equation}
  \langle 0 | \hat{\varphi}^{2}(x)| 0\rangle_{\alpha}  = 
\langle 0 | \hat{\varphi}^{2}(x)| 0\rangle_{\subN }
 - L^{-1} \cos ^{2} \rho  \cot \alpha  \int _{{\widetilde{\mathcal {I}}}_{\pi /2}} G_{\subN}^{\subESU}(x,y)G_{\alpha }^{\subESU}(y,x) \, \rmd S ,
 \label{eq:VPRobinAdSalt}
\end{equation}
where the integral is performed over points $y$ lying in the hypersurface ${\widetilde {\mathcal {I}}}_{\pi /2}$ in ESU.
We have shown that the integral diverges like $\epsilon ^{-1}$ when the point $x$ approaches the boundary, $\rho = \pi /2-\epsilon $ with $\epsilon \rightarrow 0$.
In (\ref{eq:VPRobinAdSalt}), this divergent integral is multiplied by a factor of $\cos ^{2}\rho =\epsilon ^{2}+\Or (\epsilon ^{3})$ and therefore we deduce that, on the CAdS boundary,
\begin{equation}
  \lim _{\rho \rightarrow \frac{\pi }{2}}\langle 0 | \hat{\varphi}^{2}(x)| 0\rangle_{\alpha}  = 
\lim _{\rho \rightarrow \frac{\pi }{2}} \langle 0 | \hat{\varphi}^{2}(x)| 0\rangle_{\subN }  = \frac{5}{48\pi ^{2}L^{2}}.
\end{equation}
Therefore we have shown that, for all Robin boundary conditions other than Dirichlet boundary conditions, the vacuum expectation value approaches that for Neumann boundary conditions on the CAdS boundary, in accordance with our numerical results.

The construction of \cite{Deutsch:1978sc} can also be applied to the thermal Euclidean Green's functions.
Since, like the vacuum Euclidean Green's function, the thermal Euclidean Green's function $G^{\subESU}_{\alpha , \beta }(x,x')$ satisfies the inhomogeneous scalar field equation (\ref{eq:inhomwave}) and the boundary conditions (\ref{eq:ESUboundary}), the argument leading to (\ref{eq:Deutsch1}) holds also for the thermal Euclidean Green's function, hence
\begin{equation}
    G_{\alpha ,\beta  }^{\subESU}(x,x') = G_{\subN}^{\subESU}(x,x')
    - L^{-1}\cot \alpha  \int _{{\widetilde{\mathcal {I}}}_{\pi /2}} G_{\subN}^{\subESU}(x,y)G_{\alpha ,\beta }^{\subESU}(y,x') \, \rmd S ,
    \label{eq:Deutsch2}
\end{equation}
where the integral is over the space-time points $y$.
Note that (\ref{eq:Deutsch2}) involves the vacuum Euclidean Green's function for Neumann boundary conditions.
Substituting for $G_{\alpha ,\beta  }^{\subESU}(y,x')$ in the integral on the right-hand-side gives
\begin{eqnarray}
    G_{\alpha ,\beta  }^{\subESU}(x,x') & = & G_{\subN}^{\subESU}(x,x')
    - L^{-1}\cot \alpha  \int _{{\widetilde{\mathcal {I}}}_{\pi /2}} G_{\subN}^{\subESU}(x,y)G_{\subN}^{\subESU}(y,x') \, \rmd S 
    \nonumber \\ & & 
   \hspace{-2cm} +L^{-2} \cot ^{2}\alpha  
    \int _{{\widetilde{\mathcal {I}}}_{\pi /2}} G_{\subN  }^{\subESU}(x,y)
\left[ 
\int _{{\widetilde {\mathcal {I}}}_{\pi /2}} G_{\subN }^{\subESU}(y,z)
G_{\alpha, \beta }^{\subESU}(z,x') \, \rmd S \right]  \, \rmd S ,
    \label{eq:Deutsch3}
\end{eqnarray}
where the inner integral in the final term is over the space-time points $z$.
Comparing (\ref{eq:Deutsch3}) with (\ref{eq:ESUGalphaexp}, \ref{eq:GESUiterate}), we see that, while the vacuum and thermal Euclidean Green's functions are not the same, the first two terms in their asymptotic expansions are identical.
Following the analysis of \cite{Deutsch:1978sc}, the divergences in the renormalized expectation values as the boundary is approached are due to the first two terms in (\ref{eq:ESUGalphaexp}, \ref{eq:Deutsch3}) and hence are identical for vacuum and thermal states, in accordance with the results of \cite{Kennedy:1979ar}.
This means that the vacuum and thermal Euclidean Green's functions on ESU differ by terms which are finite on the boundary ${\widetilde {\mathcal {I}}}_{\pi /2}$ in ESU. 
The above analysis for the vacuum expectation values on CAdS therefore extends trivially to the thermal expectation values to give
\begin{equation}
  \lim _{\rho \rightarrow \frac{\pi }{2}}\langle \beta  | \hat{\varphi}^{2}(x)| \beta \rangle_{\alpha}  = 
\lim _{\rho \rightarrow \frac{\pi }{2}} \langle 0 | \hat{\varphi}^{2}(x)| 0 \rangle_{\subN }  = \frac{5}{48\pi ^{2}L^{2}},
\end{equation}
again in agreement with our numerical results.

\section{Conclusions}
\label{sec:conc} 

This paper has been concerned with the renormalized vacuum polarization for a massless, conformally coupled scalar field on the (covering space of) global four-dimensional AdS. 
Robin boundary conditions, parameterized by $\alpha \in [0,\pi )$ (\ref{eq:Robin}) are applied to {\em {all}} modes of the scalar field on the space-time boundary. 
We work in the context of a semiclassical approximation to quantum gravity, where the CAdS space-time is fixed and purely classical, and a quantum scalar field propagates on this background.
In section \ref{sec:intro}, we raised five questions and we now discuss the implications for these of the results presented here.

First, question \ref{qu1} asked ``Are general Robin boundary conditions physically valid?''. 
For a classical scalar field, this question had previously been answered in \cite{Ishibashi:2004wx}, where it was shown that there are classically unstable modes for $\pi >\alpha > \alpha _{\rm {crit}}$, where $\alpha _{\rm {crit}}$ is given by (\ref{eq:alphacrit}) for a massless and conformally coupled scalar field in four space-time dimensions. 
When $0\le \alpha <  \alpha _{\rm {crit}}$, the classical evolution of the scalar field is defined consistently.
Our work has shown that, for $0\le \alpha <  \alpha _{\rm {crit}}$, quantum scalar fields  satisfying Robin boundary conditions have finite renormalized vacuum polarization. 
However, the magnitude of both vacuum and thermal expectation values diverges as $\alpha \rightarrow \alpha _{\rm {crit}}$, indicating a breakdown in the semiclassical approximation.
This approximation ignores the backreaction of the quantum field on the space-time geometry, and is hence no longer valid when the quantum fluctuations of the field are not small.

In this paper we have considered both vacuum and thermal expectation values. Question \ref{qu2} asked whether these states are Hadamard.
Providing the scalar field has no classical instabilities, in \cite{Dappiaggi:2018xvw} vacuum states on CAdS are constructed for which the Green's function has the Hadamard form.
Our numerical calculations have found finite thermal expectation values when the scalar field is classically stable.
This indicates that thermal states also have the Hadamard form, in other words the difference between the thermal and vacuum Green's functions with the same boundary conditions applied is regular in the coincidence limit. 

Question \ref{qu3} raised the important question of the symmetries satisfied by the quantum states.
When either Dirichlet or Neumann boundary conditions are applied, vacuum states respect all the symmetries of the underlying CAdS space-time, but thermal states break translation symmetry by selecting a preferred space-time point relative to which the temperature is defined \cite{Allen:1986ty, Ambrus:2018olh}.
When other Robin boundary conditions are applied to the scalar field, we have found that even  vacuum states do not possess all the symmetries of CAdS.
This is in agreement with the results of \cite{Barroso:2019cwp}, where  Robin boundary conditions were applied only to the $\ell=0$ field modes, whereas we have applied Robin boundary conditions consistently to all field modes.
Thermal states with Robin boundary conditions applied also do not have maximal symmetry.

Next, question \ref{qu4} asks how we can practically compute quantum expectation values for vacuum and thermal states. 
We began by considering expectation values defined on Lorentzian CAdS space-time.
As found in \cite{Barroso:2019cwp}, calculating these directly is computationally very challenging. 
We have therefore adopted an alternative approach by working on the Euclidean section of CAdS.
This enables the Hadamard renormalization prescription to be applied to the vacuum polarization, yielding mode sums which are amenable to numerical computation for all valid Robin boundary conditions. 

This practical methodology enabled us to address question \ref{qu5}, namely ``Does the vacuum polarization asymptote  to a finite value for arbitrary Robin boundary conditions?''.
Providing the boundary conditions are such that the scalar field has no classical instabilities, the vacuum polarization approaches a finite limit as the space-time boundary is approached.
Our numerical computations indicated that this limit is the same for all Robin boundary conditions except for Dirichlet boundary conditions.
Analysis based on the general framework in \cite{Deutsch:1978sc} showed that this is indeed the case.

While most of the literature on expectation values of a quantum scalar field on CAdS (such as \cite{Ambrus:2018olh,Kent:2014nya}) has considered only Dirichlet boundary conditions, our work shows that these are nongeneric and have rather different properties on the CAdS boundary compared with other Robin boundary conditions, including Neumann.
In this paper we have considered only the simplest expectation value, the vacuum polarization of the scalar field.
It would be very interesting to study whether the behaviour we have found, both as $\alpha \rightarrow \alpha _{\rm {crit}}$ and $\rho \rightarrow \pi /2$, extends to the renormalized expectation value of the quantum stress-energy tensor.
This expectation value governs the backreaction of the quantum field on the space-time geometry via the semiclassical Einstein equations, and therefore the possible breakdown of the semiclassical approximation as $\alpha \rightarrow \alpha _{\rm {crit}}$ (implied by our results for the vacuum polarization) could be addressed. 
The behaviour of the renormalized stress-energy tensor on the space-time boundary would also merit investigation.

Our work in this paper has focussed on a massless, conformally-coupled scalar field. 
A natural extension would be to consider the case of either a massive scalar field or nonconformal coupling. 
Calculations of renormalized expectation values for nonconformally coupled scalar fields are complicated by the fact that the Hadamard parametrix (\ref{eq:GS}) contains logarithmic singularities which are absent when the field is conformally coupled \cite{Decanini:2005eg,Kent:2014nya}. 
These additional singularities will present technical challenges for any future computation of the renormalized vacuum polarization in this case. 

Finally, in this paper we have considered pure CAdS space-time. The requirement to impose boundary conditions on a quantum scalar field applies not only to this space-time, but to all asymptotically-AdS space-times, including black holes. 
Vacuum polarization on spherically symmetric, asymptotically-AdS black holes has been computed for both massless, conformally coupled scalar fields \cite{Flachi:2008sr} and more general scalar fields \cite{Breen:2018ukd} satisfying Dirichlet boundary conditions.
Static, vacuum, asymptotically-AdS black holes are not necessarily spherically symmetric, and the vacuum polarization has also been studied for asymptotically Lifshitz black holes \cite{Quinta:2016eql} and topological black holes \cite{Morley:2018lwn}, again for Dirichlet boundary conditions.
The effect of boundary conditions on the renormalized vacuum polarization on asymptotically-AdS black holes would make for interesting future work.

\appendix

\section{Evaluation of (\ref{eq:intappendix})}

In this appendix we give details of the evaluation of the integral in (\ref{eq:intappendix}):
\begin{eqnarray}
    {\mathfrak {I}} &=& \int_{0}^{\pi/2}\tan^{2}\rho\,\chi_{n\ell}(\rho)\chi_{n'\ell}(\rho) \, \rmd\rho \nonumber \\
    \fl &=& \int_{0}^{\pi/2}\sin\rho\,\mathbf{Q}_{\omega-1/2}^{\ell+1/2}(\cos\rho)\mathbf{Q}_{\omega'-1/2}^{\ell+1/2}(\cos\rho)\, \rmd \rho \nonumber\\
    \fl &=& \frac{1}{\Gamma(\ell+\omega+1)\Gamma(\ell+\omega'+1)}\int_{0}^{1}Q_{\omega-1/2}^{\ell+1/2}(x)Q_{\omega'-1/2}^{\ell+1/2}(x) \, \rmd x.
\end{eqnarray}
From the ODE satisfied by the Legendre functions, we have 
\begin{eqnarray}
\fl & & \frac{\rmd }{\rmd x}\left[(1-x^{2})
\left(Q_{\omega'-1/2}^{\ell+1/2}(x)
\frac{\rmd Q_{\omega-1/2}^{\ell+1/2}}{\rmd x}
-Q_{\omega-1/2}^{\ell+1/2}(x)
\frac{\rmd Q_{\omega'-1/2}^{\ell+1/2}}{\rmd x}\right)\right]
\nonumber \\ & & \qquad \qquad
=(\omega'^{ \, 2}-\omega^{2})Q_{\omega-1/2}^{\ell+1/2}(x)Q_{\omega'-1/2}^{\ell+1/2}(x).    
\end{eqnarray}
Integrating both sides from $A$ to $B$, for $B>A$, gives
\begin{eqnarray}
\fl & & \hspace{-1cm}
(\omega'^{ \, 2}-\omega^{2})\int_{A}^{B}Q_{\omega-1/2}^{\ell+1/2}(x)Q_{\omega'-1/2}^{\ell+1/2}(x) \, \rmd x
\nonumber \\ & & \quad  
= (1-B^{2}) \left[ 
Q_{\omega'-1/2}^{\ell+1/2}(B)
\frac{\rmd Q_{\omega-1/2}^{\ell+1/2}}{\rmd x}(B)
-Q_{\omega-1/2}^{\ell+1/2}(B)
\frac{\rmd Q_{\omega'-1/2}^{\ell+1/2}}{\rmd x}(B) \right] 
\nonumber \\ & & \qquad 
 - (1-A^{2}) \left[ 
 Q_{\omega'-1/2}^{\ell+1/2}(A)
 \frac{\rmd Q_{\omega-1/2}^{\ell+1/2}}{\rmd x}(A)
 -Q_{\omega-1/2}^{\ell+1/2}(A)
 \frac{\rmd Q_{\omega'-1/2}^{\ell+1/2}}{\rmd x}(A) \right] .
 \nonumber \\ & & \qquad 
\end{eqnarray}
From standard properties of Legendre functions \cite{NIST:DLMF}, we find that, in the limit $B\rightarrow 1$,
\begin{equation}
    (1-B^{2})Q_{\omega-1/2}^{\ell+1/2}(B)
    \frac{\rmd Q_{\omega'-1/2}^{\ell+1/2}}{\rmd x}(B)\propto (1-B^{2})^{\ell+1/2}\rightarrow 0
\end{equation}
and so
\begin{eqnarray}
\fl & &  \hspace{-1cm}
(\omega'^{ \, 2}-\omega^{2})\int_{A}^{1}Q_{\omega-1/2}^{\ell+1/2}(x)Q_{\omega'-1/2}^{\ell+1/2}(x) \, \rmd x 
\nonumber \\ & & \quad 
=  (1-A^{2}) 
\left[ 
Q_{\omega-1/2}^{\ell+1/2}(A)
\frac{\rmd Q_{\omega'-1/2}^{\ell+1/2}}{\rmd x}(A)
- Q_{\omega'-1/2}^{\ell+1/2}(A)
\frac{\rmd Q_{\omega-1/2}^{\ell+1/2}}{\rmd x}(A) 
\right] .
\label{eq:appendix1}
\end{eqnarray}

Considering the case $\omega\neq \omega'$, on taking the limit $A\rightarrow 0$ the right-hand-side of (\ref{eq:appendix1}) is zero due to the boundary conditions (\ref{eq:Robin}). 
When $\omega'\rightarrow\omega$, we use properties of Legendre functions close to the origin \cite{NIST:DLMF} to give
\begin{eqnarray}
\fl & & 
\hspace{-2cm}
(\omega'^{ \, 2}-\omega^{2})\int_{0}^{1}Q_{\omega-1/2}^{\ell+1/2}(x)Q_{\omega'-1/2}^{\ell+1/2}(x) \, \rmd x 
\nonumber \\ & & 
\hspace {-1.7cm} = 2^{2\ell-1}\pi\sin\left(\frac{1}{2}(\omega+\omega'+2\ell)\pi\right)
\nonumber\\ & & 
\hspace{-0.5cm}
 \times\left[\frac{\Gamma(\frac{1}{2}(\omega+\ell+2))\Gamma(\frac{1}{2}(\omega'+\ell+1))}{\Gamma(\frac{1}{2}(\omega-\ell))\Gamma(\frac{1}{2}(\omega'-\ell+1))}-\frac{\Gamma(\frac{1}{2}(\omega+\ell+1))\Gamma(\frac{1}{2}(\omega'+\ell+2))}{\Gamma(\frac{1}{2}(\omega-\ell+1))\Gamma(\frac{1}{2}(\omega'-\ell))}\right]
 \nonumber\\ & & 
 \hspace {-1.5cm}
 +2^{2\ell-1}\pi\sin\left(\frac{1}{2}(\omega-\omega')\pi\right)
 \nonumber\\  & & 
\hspace{-0.5cm} \times\left[\frac{\Gamma(\frac{1}{2}(\omega+\ell+2))\Gamma(\frac{1}{2}(\omega'+\ell+1))}{\Gamma(\frac{1}{2}(\omega-\ell))\Gamma(\frac{1}{2}(\omega'-\ell+1))}+\frac{\Gamma(\frac{1}{2}(\omega+\ell+1))\Gamma(\frac{1}{2}(\omega'+\ell+2))}{\Gamma(\frac{1}{2}(\omega-\ell+1))\Gamma(\frac{1}{2}(\omega'-\ell))}\right].
 \nonumber\\  & & 
\end{eqnarray}
We now divide both sides by $(\omega'^{ \, 2}-\omega^{2})$ and evaluate limits using L'Hopital's rule to obtain
\begin{eqnarray}
\fl & & \hspace{-2cm} 
\int_{0}^{1}Q_{\omega-1/2}^{\ell+1/2}(x)Q_{\omega'-1/2}^{\ell+1/2}(x) \, \rmd x
\nonumber \\ & = &  
\frac{2^{2\ell-2}\pi\Gamma(\frac{1}{2}(\ell+\omega+1))\Gamma(\frac{1}{2}(\ell+\omega+2))}{\omega\Gamma(\frac{1}{2}(\omega-\ell+1))\Gamma(\frac{1}{2}(\omega-\ell))}
\nonumber\\
\fl & & \times \left[\sin((\omega+\ell)\pi)(-\zeta(\omega-\ell+1)-\zeta(\omega-\ell))+\pi\right]\delta_{\omega\omega'},
\fl
\end{eqnarray}
where $\zeta(z)$ is defined in (\ref{eq:zeta}). 
Noting the dependence of $\omega$ on $n$ by writing $\omega=\omega_{n\ell}$, then, using relations for Gamma functions \cite{NIST:DLMF}, we recover the result (\ref{eq:intappendix}) for the integral ${\mathfrak {I}}$:
\begin{equation}
   {\mathfrak {I}}  =  \delta_{n n'}\frac{\pi\left[\pi-\sin(\pi(\omega_{n\ell}+\ell))(\zeta(\ell+\omega_{n\ell}+1)+\zeta(\omega_{n\ell}-\ell))\right]}{8\,\omega_{n\ell}\,\Gamma(\ell+\omega_{n\ell}+1)\Gamma(\omega_{n\ell}-\ell)}.
\end{equation}

\ack
T.M.~thanks the School of Mathematics and Statistics at the University of Sheffield for the provision of a studentship supporting this work. 
The work of E.W.~is supported by the Lancaster-Manchester-Sheffield Consortium for Fundamental Physics under STFC grant ST/P000800/1 and partially supported by the H2020-MSCA-RISE-2017 Grant No.~FunFiCO-777740.

\end{document}